\documentclass[10pt,conference,dvipsnames,svgnames,x11names,table]{IEEEtran}
\usepackage{tikz}
\usepackage{fancyhdr}
\usepackage[newfloat,frozencache,cachedir=.]{minted}
\usepackage[utf8]{inputenc}
\usepackage[T1]{fontenc}
\usepackage{hyphenat} 
\PassOptionsToPackage{hyphens}{url}\usepackage[draft]{hyperref}
\usepackage[british]{babel}
\addto\captionsbritish{\renewcommand{\figurename}{Fig.}}
\usepackage{xspace}
\usepackage{ifthen}
\usepackage[inline]{enumitem}
\usepackage{amsmath,amssymb,bm}
\usepackage{graphicx}
\usepackage{array}
\usepackage{setspace}
\usepackage[abbreviations]{foreign}
\usepackage{ragged2e}
\usepackage{flushend}
\usepackage{microtype}
\usepackage[acronym]{glossaries}
\usepackage[leftmargin=\parindent,vskip=0pt]{quoting}
\usepackage{booktabs}
\usepackage{xcolor}
\usepackage{times}
\usepackage[scaled=0.85]{beramono}
\usepackage{pifont}
\usepackage{framed}
\colorlet{shadecolor}{Azure2}
\newcommand{\subparagraph}{}
\usepackage[compact]{titlesec}
\usepackage[font={small,stretch=1.0},labelfont=bf,skip=2pt,belowskip=-4pt]{caption}
\usepackage[moderate,lists=normal]{savetrees}

\titlespacing\section{8pt}{6pt plus 2pt minus 2pt}{2pt plus 2pt minus 2pt}
\titlespacing\subsection{0pt}{4pt plus 2pt minus 2pt}{2pt plus 1pt minus 1pt}
\titlespacing\paragraph{0pt}{2pt plus 0pt minus 1pt}{1.0ex}

\hypersetup{
  colorlinks,
  linkcolor={green!50!black},
  citecolor={red!70!black},
  urlcolor={blue!70!black}
}

\newcommand{\sys}{\textls[-50]{\rm\textsc{Pal{\ae}mon}}\xspace}
\newcommand{\sysrm}{P{\small AL{\AE}MON}\xspace}

\newboolean{showcomments}
\setboolean{showcomments}{true}
\ifthenelse{\boolean{showcomments}}
{ \newcommand{\mynote}[3]{
    \fbox{\bfseries\sffamily\scriptsize#1}
    {\small$\blacktriangleright$\textsf{\emph{\color{#3}{#2}}}$\;\blacktriangleleft$}}}
{ \newcommand{\mynote}[3]{}}

\newcommand{\unit}[1]{\mbox{\hspace{2pt}#1}\xspace}
\newcommand{\setup}[1]{\textsf{\small{}#1\xspace}}

\newcommand{\spaceafterfloat}{\vspace{-10pt}}

\newacronym{sgx}{SGX}{software guard extensions}
\newacronym{ml}{ML}{machine learning}
\newacronym{mre}{\texttt{MRE}}{cryptographic hash of the enclave~\cite{costan2016intel}}
\newacronym{tee}{TEE}{trusted execution environment}
\newacronym{tcb}{TCB}{trusted computing base}
\newacronym{kms}{KMS}{key management system}
\newacronym{emu}{EMU}{emulation mode}
\newacronym{epc}{EPC}{enclave page cache}
\newacronym{ip}{IP}{intellectual property}
\newacronym{hsm}{HSM}{hardware security module}
\newacronym{tls}{TLS}{transport layer security}
\newacronym{dns}{DNS}{domain name system}
\newacronym{os}{OS}{operating system}
\newacronym{ca}{CA}{certification authority}
\newacronym{rc}{RC}{root certificate}
\newacronym{pki}{PKI}{private key infrastructure}
\newacronym{kek}{KEK}{key encryption key}
\newacronym{ias}{IAS}{Intel attestation service}
\newacronym{orm}{ORM}{object-relational mapping}
\newacronym{pin}{PIN}{personal identification number}
\newacronym{sms}{SMS}{secret management service}
\newacronym{tms}{TMS}{trust management service}
\newacronym{pqe}{PQE}{\sys quoting enclave}
\newacronym{qe}{QE}{quoting enclave~\cite{costan2016intel}}
\newacronym{bft}{BFT}{Byzantine Fault Tolerance~\cite{castro1999practical}}
\newacronym{crud}{CRUD}{create, read, update and delete}
\newacronym{cif}{CIF}{confidentiality, integrity and freshness}
\newacronym{ups}{UPS}{uninterruptible power supply}
\newacronym{sdkms}{SDKMS}{self-defending key management service}

\glsdisablehyper
\makeglossaries
\makeatletter
\renewcommand{\sectionautorefname}{\S\@gobble}
\renewcommand{\subsectionautorefname}{\S\@gobble}
\makeatother
\usepackage{tcolorbox}
\title{\LARGE Trust Management as a Service:\\
	 Enabling Trusted Execution in the Face of Byzantine Stakeholders}

\author{\IEEEauthorblockN{Franz Gregor\IEEEauthorrefmark{1}, Wojciech Ozga\IEEEauthorrefmark{1}, Sébastien Vaucher\IEEEauthorrefmark{2}, Rafael Pires\IEEEauthorrefmark{2}, Do Le Quoc\IEEEauthorrefmark{1}, Sergei Arnautov\IEEEauthorrefmark{3},\\
André Martin\IEEEauthorrefmark{1}, Valerio Schiavoni\IEEEauthorrefmark{2}, Pascal Felber\IEEEauthorrefmark{2}, Christof Fetzer\IEEEauthorrefmark{1}}
\IEEEauthorblockA{TU Dresden, Germany\IEEEauthorrefmark{1} --- Université de Neuchâtel, Switzerland\IEEEauthorrefmark{2} --- Scontain UG, Germany\IEEEauthorrefmark{3} \vspace{-8pt}}
}

\begin{document}

\maketitle
\thispagestyle{fancy}

\begin{abstract}

Trust is arguably the most important challenge for critical services both deployed as well as accessed remotely over the network.
These systems are exposed to a wide diversity of threats, ranging from bugs to exploits, active attacks, rogue operators, or simply careless administrators.
To protect such applications, one needs to guarantee that they are properly configured and securely provisioned with the ``secrets'' (\eg, encryption keys) necessary to preserve not only the confidentiality, integrity and freshness of their data but also their code.
Furthermore, these secrets should not be kept under the control of a single stakeholder---which might be compromised and would represent a single point of failure---and they must be protected across software versions in the sense that attackers cannot get access to them via malicious updates.
Traditional approaches for solving these challenges often use \emph{ad hoc} techniques and ultimately rely on a \gls{hsm} as root of trust.
We propose a more powerful and generic approach to trust management that instead relies on \glspl{tee} and a set of stakeholders as root of trust.
Our system, \sys, can operate as a managed service deployed in an untrusted environment, \ie, one can delegate its operations to an untrusted cloud provider with the guarantee that  data will remain confidential despite not trusting any individual human (even with root access) nor system software.
\sys addresses in a secure, efficient and cost-effective way five main challenges faced when developing trusted networked applications and services.
Our evaluation on a range of benchmarks and real applications shows that \sys performs efficiently and can protect secrets of services without any change to their source code.

\end{abstract}
 \section{Introduction}
\label{sec:introduction}
\glsresetall

\begin{figure}[t!]
	\centering
	\includegraphics[scale=0.65]{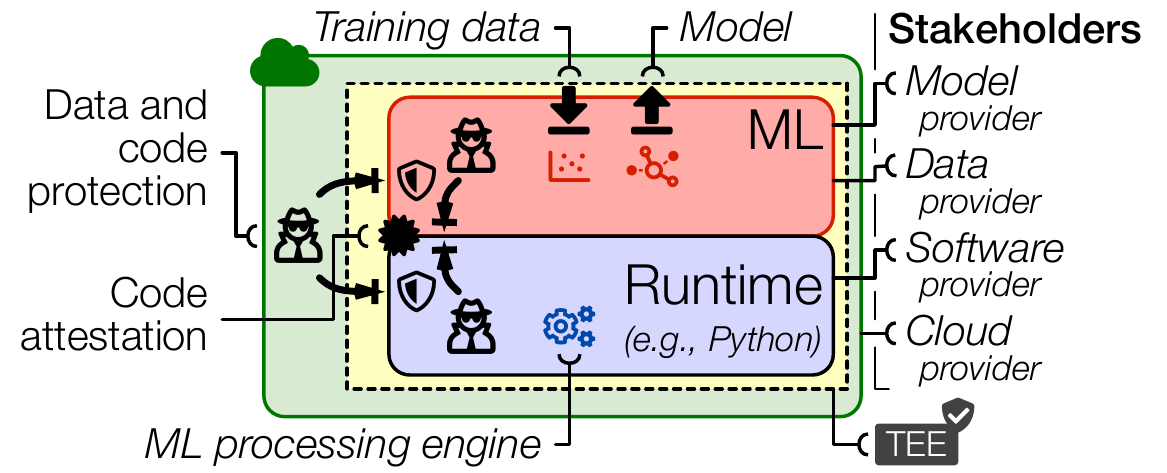}
	\caption{Components and stakeholders of the \gls{ml} use case. The training data and model belong to the data and model providers, whereas the \gls{ml} processing runtime is owned by the software provider. These key assets must be protected at all times from the other stakeholders.}
	\label{fig:ml}
\end{figure}

Protecting the \gls{cif} of application data is a key challenge of many applications, and a primary reason for companies to be wary of deploying their system outside premises in shared environments.
To illustrate the challenges faced in such scenarios, consider for instance modern machine learning applications that require significant computing power and would hence benefit from running in a scalable cloud infrastructure.
Yet, at the same time, they need to protect their code, their training data (input) and the produced model (output), all of which represent key assets for their respective owners (see \autoref{fig:ml}).
The development and operation of such a large system involve multiple stakeholders, notably software developers, system administrators, data providers and cloud providers, which cannot necessarily be trusted and might collude to gain advantages over the other stakeholders~\cite{noor2013trust}.
For example, we cannot trust that system administrators or software developers will neither leak~\cite{win8tradesecret, amazontradesecret} nor modify application code and data.
To address these challenges and enable \emph{trusted application execution in the face of Byzantine stakeholders}, we have designed \sys, a trust management service that builds on top of the SCONE platform~\cite{arnautov2016scone} and ultimately relies on hardware-based \glspl{tee} for secure execution.

\sys was originally developed in order to address real problems from some users and there was no alternative service available to solve the problems we faced.
It was extended and refined over the course of the last two years to take into account additional threats and support the evolution of trusted applications via secure updates.
We motivate its design and illustrate its operation on a real-life production use case in the machine learning space, yet our approach is more general and applies to a much wider range of applications.

At the core of \sys is a \emph{trust management service} specifically designed to address the following main challenges:
\setlist{nolistsep}
\begin{enumerate}[leftmargin=*,noitemsep]
\item \textbf{Secret management ---} How can we securely provide applications with secrets in an untrusted environment?
\item \textbf{Managed operation ---} How can we delegate the management of \sys to untrusted stakeholders?
\item \textbf{Robust root of trust ---} How can we protect \gls{cif} against malicious stakeholders?
\item \textbf{Rollback protection ---} How can we ensure freshness of data and code in an efficient manner?
\item \textbf{Secure update ---} How can we support secure updates of applications and \sys?
\end{enumerate}

To better understand the importance of these challenges, let us illustrate them on one of the original use cases that guided the design of \sys.
This real-life use case comes from a company, a \emph{software provider} specialized in \gls{ml} that develops its \gls{ml} engine in Python (see \autoref{fig:ml}).
The engine is executed by a second party, the \emph{model provider}, which processes training data to produce a model.
The software provider must neither learn the training data nor the produced model.
Conversely, the model provider must not learn the application code of the software provider.
Moreover, the software provider may want to limit the number of models produced by its application, and hence the number of times the code is executed.
The model provider might try to circumvent this limitation by reverting the application to a previous state in order to generate more models (``rollback attack'').

In its initial deployment, the model provider ran the ML engine on dedicated computers within its own air-gapped infrastructure, over which it has complete control.
Then, the model provider wants to execute the application in a cloud instead.
Therefore, it also needs to protect its training data and generated models from the third-party cloud provider, including developers and system administrators.
Finally, when operating in the untrusted cloud, it should be possible for the software provider as well as the provider of the Python runtime to continuously update their software while preserving the trust guarantees, \ie, one should prevent an attacker from injecting malicious code during software updates.

This real-life application illustrates the need for addressing the aforementioned set of problems: it requires secure \emph{secret management} to preserve confidentiality of the code and data, \emph{managed operation} to delegate the management to a cloud provider, a \emph{robust root of trust} to protect both
data and code, \emph{rollback protection} to control the number of models produced, and support for \emph{secure update} for managing the life cycle of software deployed in a third-party cloud.
Our general approach to address these five problems is to define a novel \gls{tms} that supports security policies and is able to deal with untrusted stakeholders.
The \gls{tms} can itself be managed by an untrusted entity (\eg, the \gls{ml} model provider) while still being trusted by other entities (\eg, the \gls{ml} software provider).

The contributions of this paper are as follows.
While many of the techniques that we use are known, combining these to address the problems we face is novel.
We are not aware of any other service that transparently protects applications from rollback attacks with little overheads, supports secure software updates and guarantees \gls{cif} of data and code even in the face of insider attacks, while still being able to delegate the management of the service to remote providers. Throughput of our monotonic counters is 5 orders of magnitude higher than those provided by the SGX platform.

The remainder of the paper is organized as follows.
We first introduce the threat model and problem in \autoref{sec:problem}.
We then present the architecture of \sys in \autoref{sec:approach} and its implementation in \autoref{sec:implementation}.
We evaluate \sys's security and performance in \autoref{sec:evaluation}, present a real use case in production settings in \autoref{sec:usecase}, discuss related work in \autoref{sec:related} and conclude in \autoref{sec:conclusion}.

 \section{Threats and Challenges}
\label{sec:problem}

We first introduce our threat model before we describe in greater details the challenges addressed by \sys.

\subsection{Threat Model}
\label{subsec:threat-model}

Services executing in untrusted environments such as clouds are vulnerable to attackers with root privileges.
Attackers often target the credentials of system administrators to gain access to hosts~\cite{NSAKeynote,NSAhuntSA,Snowden}.
They also exploit bugs in the system software to gain root privileges on systems~\cite{Rootin5secs,KCook}.

Multiple stakeholders of the same services cannot trust each other to protect the \gls{cif} of their digital assets such as data and code.
\autoref{fig:approach} illustrates the security goals of \sys with regard to the stakeholders as well as the building blocks in our system.
For example, \sys can protect \gls{cif} of Python code deployed at remote sites, a somewhat surprising but popular requirement for several applications of our users.
Furthermore, applications must be regularly updated, and we need to protect them against malicious software updates triggered by attackers.

Note that we cannot trust any system administrator or software developer---actually not any single individual.
Hence, we do not trust in \gls{cif} of main memory.
We also do not assume that the OS-based access control can ensure \gls{cif} since a single malicious system administrator could break this assumption.
It follows that one cannot trust software updates originating from a single developer nor security policies defined by any individual, independent of their authorization level or trustworthiness.

\sys protects from attempts to compromise \gls{cif} of code and data by requiring a quorum of trusted entities---not just a single individual---to approve any given change.
This is based on the assumption that, in any organisation that must securely operate an application, one can identify a set of $n$~stakeholders and a threshold~$f$ (with $f\!{<}n$), such that $n{-}\!f$ stakeholders can be trusted at any point in time, \ie, at most $f$ of them exhibit Byzantine behaviour because of neglect or malicious intent.
Hence, if at least $f\!{+}1$ stakeholders approve a change, such as a software update, at least one of them judged it to be trustworthy.
In practice, the typical convention is that any policy change must be approved by all members, \ie, stakeholders in a legal business contract.
A single member can choose to decline and in this way prevent malicious policy changes.

\glspl{tee} such as Intel \gls{sgx} are typically vulnerable to side-channel attacks~\cite{brasser2017software,lee2017inferring,chen2018sgxpectre,vanbulck2018foreshadow,weisse2018foreshadowNG}.
Such attacks can be addressed using existing techniques (\eg, \emph{Varys}~\cite{oleksenko2018varys}) and are out of the scope of this work.
Similarly, we do not consider denial of service attacks.

In our threat model, we anticipate that new attacks on \glspl{tee} can appear in the future.
We assume that we can put mitigation measures either in software or in microcode, or by limiting execution to certain CPU types and features not vulnerable to these attacks.
This implies that we need to be able to continuously update \sys as well as the applications and deactivate vulnerable instances within a short period of time, so that the system is protected against new attacks.

\subsection{Problem Statement}

We now introduce a more detailed definition of the problem, along with five challenges identified when analysing and operating a wide range of real-world applications that we briefly introduced earlier, with regard to our \gls{ml} use case.

\begin{figure}[t!]
	\centering
	\includegraphics[scale=0.65]{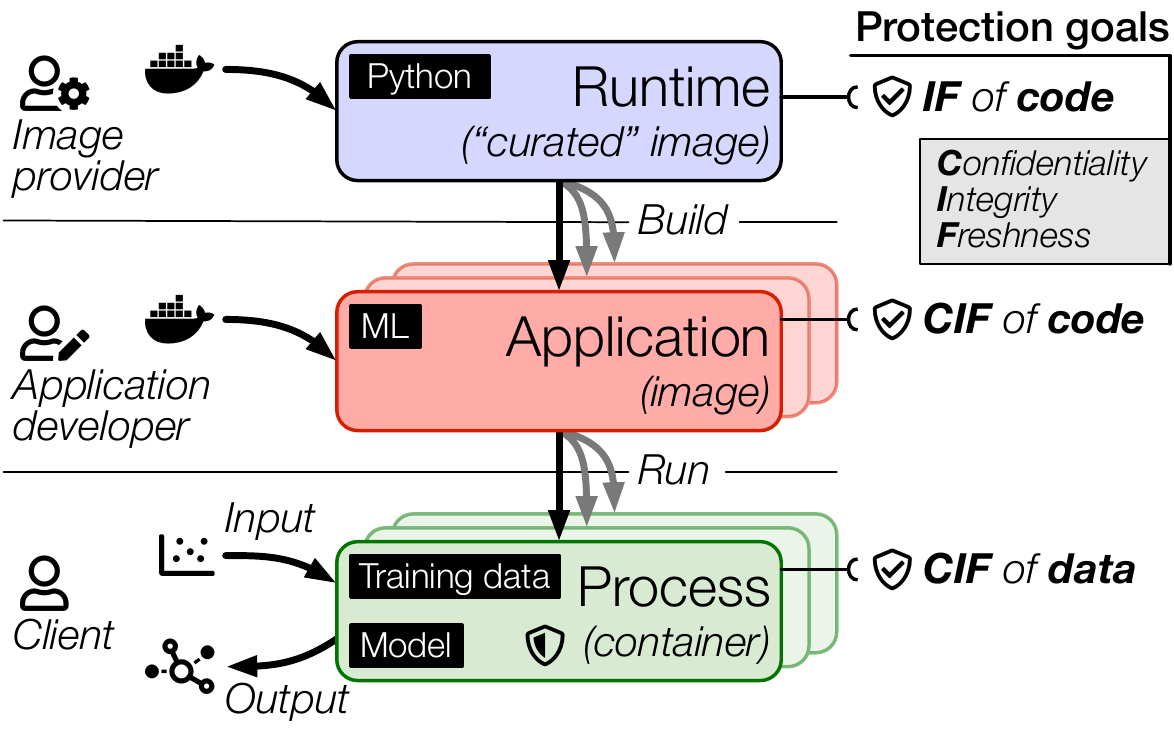}
	\caption{Stakeholders, components and protection objectives of \sys. White on black text refers to the motivating example of the machine learning application.}
	\label{fig:approach}
\end{figure}

\smallskip\noindent\textbf{Secret management ---}
Legacy software can use program arguments, environment variables or files to obtain secrets, as can be observed in~\autoref{tab:apps}, which shows a quick analysis of various popular services.
To account for this diversity and provide seamless integration of secret management in legacy applications, we need to solve the problem of:
\begin{shaded*}
\begin{quoting}
\noindent How to support secret management for common configuration approaches in a secure way and without requiring modifications to the source code?
\end{quoting}
\end{shaded*}
Container images (e.g., Docker images) are a popular way to deploy applications. We want to be able to customize container images such that not only
\begin{enumerate*}[label=\emph{(\roman*)}]
\item different application developers can inject different secrets in their derived application images, but also
\item one can inject different secrets in each container instance of an image (see \autoref{fig:approach}).
\end{enumerate*}
For example, a client running such an image might inject client-specific secrets for the application to be able to decrypt client-encrypted input files.
The injection mechanism must protect \gls{cif} in the sense that an adversary cannot read, modify or replace these secrets.

\newcommand{\YES}{\textcolor{Green4}{\ding{51}}}
\newcommand{\NO}{\textcolor{Red4}{\ding{55}}} 
\begin{table}[t!]
\scriptsize
\caption{How popular services obtain secrets ($^\ast$: evaluated in \autoref{sec:evaluation}).\vspace{-10pt}}
\label{tab:apps}
\small
\setlength{\tabcolsep}{3pt}
\center{}
\rowcolors{1}{gray!10}{gray!0}
\begin{tabular}{c>{\centering\arraybackslash}p{0.2\columnwidth}>{\centering\arraybackslash}p{0.15\columnwidth}>{\centering\arraybackslash}p{0.1\columnwidth}>{\centering\arraybackslash}p{0.1\columnwidth}>{\centering\arraybackslash}p{0.1\columnwidth}>{\centering\arraybackslash}p{0.1\columnwidth}}
  \rowcolor{gray!25}
  \textbf{Program} & Version & Lang. & Args. & Env. & Files \\
\hline
Consul
    & 1.2.3
    & Go
    &\NO
    &\YES
    &\YES
    \\
MariaDB$^\ast$
    & 10.1.26
    & C/C++
    &\YES
    &\YES
    &\YES
    \\
Memcached$^\ast$
    & 1.5.6
    & C
    &\NO
    &\NO
    &\NO
    \\
MongoDB
    & 4.0
    & C++
    &\YES
    &\YES
    &\YES
    \\
Nginx$^\ast$
    & 2.4
    & C
    &\YES
    &\YES
    &\YES
    \\
PostgreSQL
    & 10.5
    & C
    &\YES
    &\YES
    &\YES
    \\
Redis
    & 4.0.11
    & C
    &\NO
    &\NO
    &\YES
    \\
Vault$^\ast$
    & 0.8.1
    & Go
    &\YES
    &\NO
    &\YES
    \\
WordPress
    & 4.9.x
    & PHP
    &\NO
    &\NO
    &\YES
    \\
ZooKeeper$^\ast$
    & 3.4.11
    & Java
    &\NO
    &\NO
    &\YES
    \\
\hline

\hline
\end{tabular}
\spaceafterfloat
\end{table}
 
\smallskip\noindent\textbf{Managed operation ---}
The behaviour of applications is not only determined by the application code, but also by its configuration parameters such as configuration files.
Some applications could be configured in a way that can leak confidential data.
Therefore, clients would need the ability to verify that an application is properly configured to ensure \gls{cif} of their data.
This challenge is especially critical for \sys, as it manages secrets on behalf of clients while operating in a different administrative domain.
Therefore, in this paper, we address the problem of:
\begin{shaded*}
\begin{quoting}
\noindent How to delegate the management of applications, as well as \sys instances, to untrusted providers while still ensuring \gls{cif} of the secrets?
\end{quoting}
\end{shaded*}

\smallskip\noindent\textbf{Robust root of trust ---}
In addition to cloud providers and system administrators, we do not trust insiders such as the software developers who build the software components, or the security experts who design the security policies.   
We must therefore solve the problem of:
\begin{shaded*}
\begin{quoting}
\noindent How to guarantee \gls{cif} of data and code even in face of malicious insiders, \ie, in a Byzantine environment?
\end{quoting}
\end{shaded*}

\smallskip\noindent\textbf{Rollback protection ---}
Whereas cryptography can preserve confidentiality and integrity of data via encryption, it does not protect from the powerful class of ``rollback'' attacks by which a malicious party attempts to replace the current state of the file system with a previous version.
In this way, they can revert data and undo some processing.
In our \gls{ml} use case, a client could roll back the file system to execute the application more often than permitted.
Preventing rollbacks typically implies significant runtime overheads and application reengineering, which we want to avoid.
The problem we address is hence:
\begin{shaded*}
\begin{quoting}
\noindent How to protect applications from rollback attacks with only negligible overhead and without requiring modifications to the source code?
\end{quoting}
\end{shaded*}

\smallskip\noindent\textbf{Secure update ---}
Software needs to be updated continuously, not only for adding new features but more importantly to fix bugs and patch security vulnerabilities.
We therefore need a secure approach to update applications, by making sure that the new versions are genuine before transferring the secrets of the previous version to the new one.
Specifically: 
\begin{shaded*}
\begin{quoting}
\noindent How to update applications, as well as \sys itself, without compromising secrets even when facing a malicious software update initiated by an insider?
\end{quoting}
\end{shaded*}
Moreover, this should be supported in settings where the management of \sys is delegated to an untrusted party that is permitted to perform the update.

 \section{Approach: A Trust Management Service}
\label{sec:approach}
Here we describe how \sys tackles the introduced problems above, with more technical descriptions also given in \autoref{sec:implementation}.

\begin{figure*}[t!]
    \centering
    \begin{minipage}[t]{.32\textwidth}
    	\centering
		\includegraphics[scale=0.7]{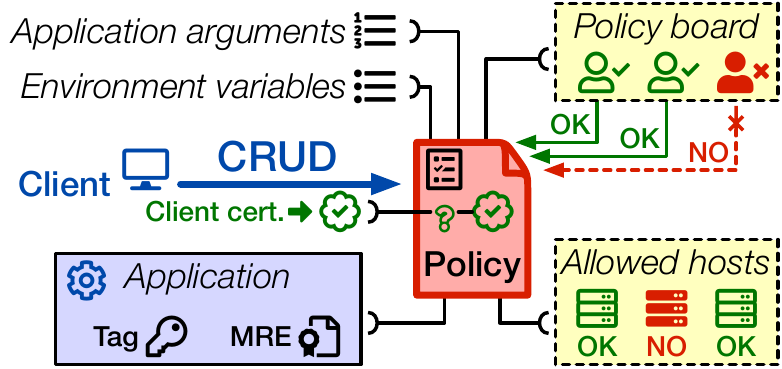}
		\caption{A policy defines which applications can access which secrets
		on which hosts.}
		\label{fig:policy}
    \end{minipage}\hfill\vline\hfill
    \begin{minipage}[t]{.32\textwidth}
		\centering
		\includegraphics[scale=0.7]{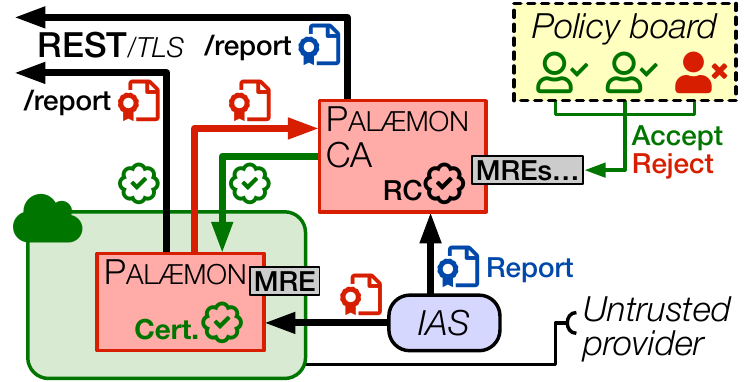}
		\caption{Managed \sys supports certifi\-cate-based attestation via \sys CA.}
		\label{fig:managed}
    \end{minipage}\hfill\vline\hfill
    \begin{minipage}[t]{.32\textwidth}
		\centering
		\includegraphics[scale=0.7]{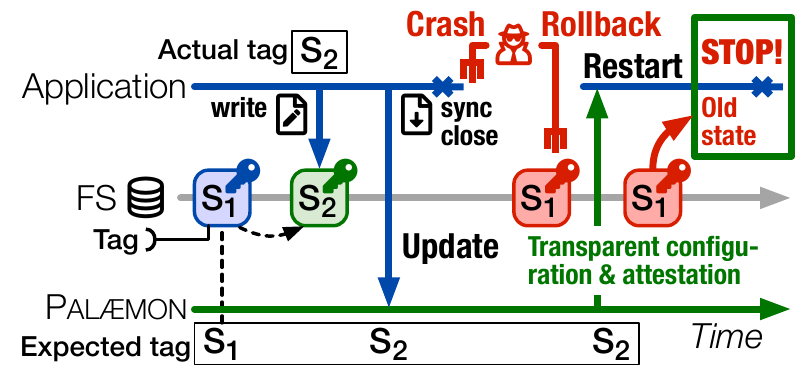}
		\caption{Application rollbacks are detected by maintaining \emph{expected tags} at \sys.}
		\label{fig:rollback}
    \end{minipage}
\end{figure*}

\subsection{Secret Management}

One of the main roles of \sys is to pass secrets in a trusted manner to applications after attesting them.
Each application is executed in a \gls{tee} and associated with a \emph{security policy} that defines \emph{which applications can access which secrets on which hosts}.
Applications are identified by a \gls{mre} and the content of the files they can access.
Secrets are typed and can either be explicitly defined, or randomly chosen by \sys.

\begin{minted}[mathescape,
linenos,
gobble=2,
frame=lines,
numbers=right,
xrightmargin=15pt,
fontsize=\small,
framesep=2mm]{yaml}
                List 1: A Palaemon Policy Example
{
		name: python_policy
		services:
		     - name: python_app
		     image_name: python_image
		     command: python /app.py  -o /encrypted-output
		     mrenclaves: ["$PYTHON_MRENCLAVE"]
		     platforms: ["$PLATFORM_ID"]
		     pwd: /
		     fspf_path: /fspf.pb
		     fspf_key: "$PALAEMON_FSPF_KEY"
		     fspf_tag: "$PALAEMON_FSPF_TAG"
		images:
		     - name: python_image
		     volumes:
		         - name: encrypted_output_volume
		         path: /encrypted-output
		volumes:
		    # an encrypted volume will 
		    # be automatically generated 
		    - name: encrypted_output_volume
		    # export encrypted volume to output policy
		    export: output_policy 
}	    
\end{minted}

Access to a security policy is guarded by a two-stage access control mechanism using a \emph{certificate} and a \emph{policy board} (see \autoref{fig:policy}). 
One can define the access control and security policy in such a way that only applications under the control of the security policy can gain access to the secrets.
In this way, one can prevent any stakeholder from accessing the secrets. 

Secrets can be passed to applications as command line arguments, environment variables, or can be injected into files. 
The files can contain \sys variables referring to the names of secrets defined in the security policy. 
The variables are transparently replaced by the value of the secret when an application that is permitted to access the secrets reads the file.  
By \emph{transparently}, we mean that the application is not aware of the replacement and its code does not need to be modified.

Secret management is supported through \emph{security policies}, whose general structure is shown in \autoref{fig:policy}.
An example of a \sys policy defined for a python application is presented in List 1.
Each policy has a unique name and can define:
\begin{enumerate*}[label=\emph{(\alph*)}]
\item the permitted \gls{mre} (line 8) of an application (several \glspl{mre} can be specified to facilitate software updates);
\item the set of permitted platforms (line 9) on which the application is permitted to run, or none if permitted to run on any platform;
\item the \emph{key} (line 12) and \emph{tag} (line 13) of the file system (the \emph{tag} is a secure hash across all files, which are transparently en/decrypted with the \emph{key} inside the \gls{tee}); 
\item the command line arguments (line 7);
\item the environment variables;
\item a set of files to inject secrets into; and
\item imports/exports of secrets from/to other policies (line 24).
\end{enumerate*}

\subsection{Managed \sysrm}

Our objective is to support a feature that we can delegate the management of a \sys instance to an untrusted party, say a cloud provider, while the clients of \sys can still trust that their secrets are safe and well protected.
Note that the cloud provider has full control over what code it executes and might try to run variants of \sys that are wrongly configured or have modified code.
We ensure that clients connecting to a \sys instance can attest it, \ie, they can verify that this instance runs the expected unmodified \sys code.
Moreover, this code does not support any configuration options that negatively influence the \gls{cif} of client data stored in the instance.

We support two ways to attest a \sys instance (see \autoref{fig:managed}): 
\begin{enumerate*}[label=\emph{(\roman*)}]
\item using \gls{tls}~\cite{Goldman:2006:LRA:1179474.1179481,6681007}; and
\item \label{item:explicit} with explicit attestation.
\end{enumerate*} 
The \gls{tls}-based attestation requires a trusted \gls{ca} with a known \gls{rc}.
The \gls{ca} first attests the \sys instance using approach \ref{item:explicit} to ensure 
that this instance runs inside a \gls{tee} and has a correct \gls{mre}.
Only then will the \gls{ca} provide the instance with a certificate signed with the \gls{rc}.
The \gls{ca} itself runs inside of a \gls{tee} and can be attested using explicit attestation. 
Entities that trust the \gls{ca} can attest the instance by checking that its \gls{tls} certificate is signed by the \gls{rc}.

To support software updates of \sys itself, the \gls{ca} includes a set of correct \glspl{mre}.
The \gls{ca} only signs certificates for these \glspl{mre} and also limits the duration of the certificates to ensure timely upgrades to new versions of \sys.
The set of \glspl{mre} is stored inside of the \gls{ca}'s binary, \ie, an adversary cannot modify the set without invalidating the \gls{mre} of the \gls{ca}.
Hence, deploying a new version of \sys requires first to deploy a new version of the \gls{ca}.
Updates of the \gls{ca} itself are controlled by a \sys policy board consisting of a set of stakeholders and follow the procedure described in \autoref{sec:swupdate}. 

Clients might not trust the \gls{ca} if they do not use the current set of valid \glspl{mre}, \eg, they only trust code instances that have been deployed some time ago, or are not represented in the \sys policy board.
These clients need to attest the \sys instance in the same way that the \gls{ca} attests instances, as described in \autoref{sec:attestation}. In practice, any updates of \sysrm must be approved by all stakeholders.

\subsection{Robust Root of Trust} 

Our threat model permits Byzantine behaviour of stakeholders like software developers and system administrators.
Any change to an application or its configuration can impact the \gls{cif} of both data and code.
\sys therefore includes a mechanism to ensure that any security policy modification must be approved by at least $f\!{+}1$ stakeholders.
To that end, a securities policy can define a \emph{policy board} and a threshold---typically set to $f\!{+}1$---of policy board members that must give approval for \sys to permit any \gls{crud} access to the policy.
Upon creation, the board of the new policy must also approve the operation.
In that way, any client can create policies as long as they have unique names, and the policy board agrees to take control over them.

Each policy board member is represented in the security policy by a \emph{certificate} and a URL of an \emph{approval service}, responsible to \emph{approve} or \emph{reject} accesses to the policy.
Upon a client access, \sys contacts the board members, verifies their certificates and asks them for approval of the request via a \gls{tls}-secured REST call to their approval service.

Approval services typically run inside \glspl{tee}.
In case the associated board member is a person, they should perform a two-factor authentication with one being based on biometric identifiers.
Approval services may also consist of services that check certain aspects of a policy, \eg, through source code analysis and verification of the \gls{mre}.
In particular, a policy board member could be an organisation that validates software, \ie, perform checks on behalf of their clients to ensure that the software associated with a certain \gls{mre} can be trusted to protect the CIF of data.

Some policy board members can be given \emph{veto} rights, \ie, they can unilaterally reject a policy change.
For example, a data provider might only provide data to applications for which it is a policy member with veto rights.
In that way, the data provider can ensure that policy changes will not result in data leakage.

\subsection{Rollback Protection}
\label{subsec:app-rollback-protection}

We want to protect applications from rollback attacks without requiring their code to be changed.
To that end, \sys performs transparent encryption of files inside of the \gls{tee}.
It uses a Merkle tree to verify integrity of the files and stores its root hash in the so-called \emph{tag} of the file system.\footnote{In reality, \sys can associate an application with multiple tags to simplify the mapping of encrypted volumes into containers.}
Any change of a file will result in a new tag, hence attempts to modify the content of the file system or to roll back to an older version can be detected by comparing the \emph{expected tag} with the \emph{actual tag} of the file system.
\sys ensures that there is no violation of integrity or freshness by verifying the value of tag on each file system access.

In order to prevent rollback attacks, it is critical to keep the {\em expected tag} value up-to-date.
This value is kept inside of a \gls{tee} but it is lost upon crash or when the application terminates (see \autoref{fig:rollback}).
Our approach is to persist the expected tag: each time
\begin{enumerate*}[label=\emph{(\roman*)}]
\item a file is closed;
\item the file system is synchronised;
\item or the application exits,
\end{enumerate*}
the runtime system pushes the \emph{expected tag} to \sys via the \gls{tls} connection that was established during application startup to perform attestation. 
\sys stores the expected values in its database.
As these values are essential for protecting the integrity and freshness of files, the expected tags must themselves be protected against rollback attacks.
We show in \autoref{sec:implementation} how \sys efficiently protects its database against rollbacks.

A policy can also define a \emph{strict mode} for an application.
In that case, \sys only permits a restart of the application if the expected tag was properly sent upon \emph{exit} during the last execution of the application.
Otherwise, the restart requires an explicit update of the policy, which is needed to adjust the tag and must in turn be approved by the policy board.  

\subsection{Secure Update}
\label{sec:swupdate}

\sys protects the \gls{cif} of both code and data.
The binary code that is initially loaded in the \gls{tee} is just \emph{integrity-} and \emph{freshness-}protected since we can only get secrets after the initial attestation of the code.
In contrast, all code that is loaded after the start of the \gls{tee} is \gls{cif}-protected.
For example, an application can load dynamic libraries in main memory, with these being transparently decrypted and \gls{cif}-protected by \sys.
Code that is loaded by interpreters and just-in-time engines is \gls{cif}-protected in the same way.

Along the same lines, \sys can also perform secure software updates of an application with the help of a policy update.
Applications are typically packaged in a container image and data (\eg, a database) is mapped into the container via a volume.
A new version of the code results in a new \gls{mre} and \emph{tag} of the container file system.
The new \gls{mre} and tag must be updated within the security policy to permit the new version to start, and this update must be approved by the policy board.

Consider the example of an image provider who maintains an image that is regularly updated, for example a Python interpreter running inside of a \gls{tee}.
As software is updated, old versions of the image should be disabled and new versions enabled.
To reduce the effort for applications that build upon this image, the provider will create a security policy defining in our example the \gls{mre} for the Python interpreter and a tag covering all the dynamic and Python libraries.
This information is exported, and can then be imported by other security policies.
Any application that uses the original image can use the exported information in its own security policy.
Additionally, the application's policy can limit the permitted combination of \glspl{mre} and tags, \eg, only allow combinations that were checked by an external service.
The application will only run with combinations that are permitted by both the image's and the application's policies.
The advantage of computing this intersection is that, if the image provider removes a combination that has become unsafe, \eg, after discovering a vulnerability, the combination will be automatically disallowed by the application's policy as well.
 \section{Implementation}
\label{sec:implementation}

In this section, we describe how to address some of the challenges we faced when implementing our approach.
Our implementation is based on the SCONE platform~\cite{arnautov2016scone} running on top of Intel \gls{sgx}~\cite{costan2016intel}.
However, note that \sysrm is designed in a generic way that can be used not only for SCONE but also for other SGX platforms such as Graphene.
We selected SCONE since it is easy to use compared to other platforms.
To run an application with Intel SGX, we just need to compile its source code with the SCONE compiler, or just link the binary of the application with the SCONE libc. 

We also considered ARM's TrustZone \gls{tee}~\cite{arm-trustzone}, but it only supports a single secure zone rather than multiple enclaves and it lacks an attestation protocol. Meanwhile, the current version of AMD's \gls{tee}, SME/SEV, lacks integrity protection and is vulnerable to server-side rollback attacks~\cite{morbitzer2018severed, hetzelt2017security, du2017secure}.
\sys runs inside a \gls{tee}, \ie, inside an \gls{sgx} enclave.
It is implemented in Rust~\cite{Matsakis:2014:RL:2692956.2663188} to ensure strong type safety. We use an encrypted embedded SQLite~\cite{owens2010sqlite} database running inside the same enclave as \sys.
We describe below how this database is protected against rollbacks without introducing any major performance bottlenecks.
\subsection{Application Attestation and Configuration}

Upon startup, an application is transparently linked with the SCONE runtime and loaded inside a \gls{tee}.
The runtime first attests the application with the help of \sys before passing control to the application.
To do so, it creates a random key pair and gets a \emph{report} from a local {\em quoting enclave}~\cite{johnson2016intel} that associates the public key with its \gls{mre}. The runtime sends the report via a newly-established \gls{tls} connection to \sys and passes along the name of its security policy which is stored in an unprotected environment variable.
The \sys instance verifies that:
\begin{enumerate*}[label=\emph{(\roman*)}]
\item the public key of the \gls{tls} client certificate matches the public key of the report;
\item the security policy name exists and the \gls{mre} is valid for the application; and
\item the application runs on a permitted platform---which we can verify with the report.
\end{enumerate*}
If this attestation succeeds, \sys sends the following data to the application: the command line arguments; the environment variables; the keys and tags for the file system; and the set of files in which secrets should be injected together with the secrets as key/value pairs.

The \sys runtime supports transparent injection of secrets into existing configuration files via a simple variable replacement mechanism.
This allows us to inject different secrets into different instances of the same application image, without the need to change the source code.
Like all files, they can be \gls{cif}-protected via transparent encryption by the \sys runtime.
The runtime injects the secrets it received from the \sys instance in each file as follows.
The file is first read in \gls{tee} memory, then parsed, and all variables found are replaced by their values.
Whenever the file is accessed, it is served from memory.
While sizeable files can also be stored encrypted in main memory or on the file system, configuration files are typically small, so we keep them in \gls{tee} memory as long as they fit.

\subsection{\sysrm Attestation}
\label{sec:attestation}

A client of a managed \sys instance must be able to ensure that the code of \sys was not modified and indeed runs inside of a \gls{tee}.
As a matter of fact, we must guarantee that an infrastructure provider cannot configure \sys in any way that breaks the trust given by the client in \sys.
We enforce this by designing \sys for its behaviour to depend solely on \gls{mre}, \ie, \sys has zero configuration parameters that affect its behaviour with regard to ensuring the \gls{cif} of the data stored in the instance by the clients.

A client connecting to a \sys instance has to attest the instance before performing any action, such as creating a new security policy.
During the initial startup, a \sys instance creates a unique public/private key pair, as well as a random key to encrypt its file system, and stores these keys in sealed storage~\cite{hoekstra2013using}.
During a restart (after an exit or a failure), the instance reads the keys from sealed storage to be able to authenticate itself.
We actually use SGX to enforce that only \sys instances on the same platform can read the sealed file.
We show in \autoref{sec:singleinstance} that at most one \sys instance at a time can start up using these keys. To handle migration issues in clouds, we can make use of existing techniques (\eg,~\cite{sgx-migration}), however, it is out of the scope of this work.

When the \sys instance starts up, it attests itself via \gls{ias}~\cite{anati2013innovative,IASv3Spec}.
On a successful attestation, it gets a report from \gls{ias} that associates its \gls{mre} with its public key.
The instance can send this report to the \sys \gls{ca} to obtain a certificate for the public key mentioned in the report.
Clients that connect to the instance via \gls{tls} are served this certificate after successful verification of the certificate by the instance.
The clients can then verify the instance via \gls{tls} by ensuring that the certificate is signed by the \sys \gls{ca}.
Alternatively, clients can request the \gls{ias} report via a REST API provided by \sys, and then verify that the report:
\begin{enumerate*}[label=\emph{(\roman*)}]
\item was indeed signed by \gls{ias}, and
\item associates the \sys \gls{mre} with the public key of the certificate.
\end{enumerate*}
At that point, the clients know that the instance runs inside of a \gls{tee} and has the correct \gls{mre}, \ie, they can now safely send requests.

Note that clients might themselves run inside a \gls{tee} and obtain the permitted \glspl{mre} from their security policy.
Moreover, they might be limited to connecting only to certain \sys instances identified by their public keys.

\subsection{Single Instance Enforcement}
\label{sec:singleinstance}

Each \sys instance is identified by a unique public key that it created during initial setup.
Upon restart, the instance will get the same public key from sealed storage.
An attacker might try to start two or more instances with the same public key and, in this way, may try to roll back some file system updates of applications.
Indeed, an application that stored its \emph{tag} in one \sys instance might be served an old tag if it is connected to the other instance after a restart.
We prevent this attack by ensuring, as part of the rollback protection (see below), that at most one instance of \sys runs with a given public key.

\subsection{Protection against Rollback Attacks}

\begin{figure}[t!]
	\centering
	\includegraphics[scale=0.65]{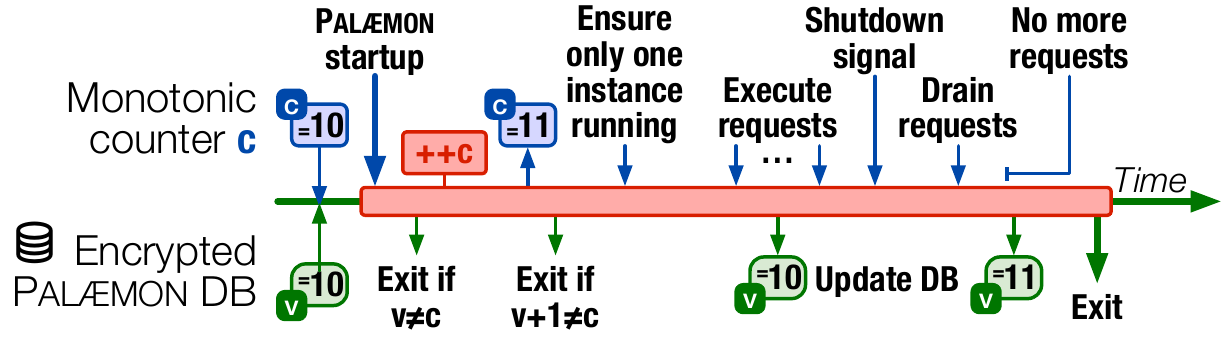}
	\caption{Rollback protection in \sys using monotonic counters.}
	\label{fig:monotonic}
\end{figure}

We use a simple yet effective approach to protect against rollback attacks on \sys's database.
The problems that we need to address are as follows.
First, the ``monotonic counters'' provided by the \gls{sgx} platform \cite{SGXMonotonicCounters} can be incremented no more than 20 times per second, \ie, at most every 50\unit{ms}.
This means that waiting for a new increment of the counter can cause delays of about 75\unit{ms}, \ie, about 25\unit{ms} on average for the current increment to finish in addition to the 50\unit{ms} of the next increment.
This delay might cause unwanted delays in applications since this limits the number of updates to the tags of a single application to about 13 times per second, which is far too few for many applications.
Moreover, continuous increment of monotonic counters introduces a wear and tear of the counters, leading to a dramatically reduced life expectancy of these counters~\cite{SGXMonotonicCounters}. 

We therefore adopt an alternative approach based on the observation that \sys runs on well-maintained hosts that have very limited unscheduled downtimes.
For example, there would typically be an \gls{ups} system to reduce the likelihood of power outages.
For any unscheduled outage, we expect that we need to perform a fail-over to another \sys service instance anyhow.
In our approach, illustrated in \autoref{fig:monotonic}, we protect against rollbacks using a version number $v$ stored in \sys's encrypted database and a hardware-based monotonic counter $c$ that keeps track of this version number.
Upon startup, \sys checks that the monotonic counter $c$ and the version $v$ of the database match, \ie, $v=c$, and otherwise exits.
\sys then increments the monotonic counter before accepting any request.
The database is now trailing the monotonic counter, \ie, $v<c$.
This will prevent any further restarts unless \sys updates $v$ during shutdown.
Furthermore, \sys checks that the increment effectively yields $c=v+1$.

Any higher value for $c$ would indicate that a second instance is already running.
In such a case, \sys would exit immediately.
As common for containers, \sys is terminated via a signal.
In that case, it shuts down all connections and stops accepting new requests.
Existing requests are still processed and the internal database is updated.
The final step is to increment the version in the database and shut down the service.
In this way, the monotonic counter and the version of the database agree again, thus allowing \sys to restart.

Note that the monotonic counter $c$ and the version $v$ are not incremented for every update to the expected tag (see $\S$\ref{subsec:app-rollback-protection}).
Thus, our rollback protection mechanism can achieve significantly higher throughput and lower latency compared to previous approaches (see $\S$\ref{sec:evaluation}).
However, in this work, we treat a systems crash as a case of attacks, \ie, we ensure the consistency and freshness with an assumption about availability.
Ensuring both consistency and availability is a challenging task, which we currently address in our ongoing work.

\subsection{Policy Access Control}

Security policies can be accessed via a REST API protected by \gls{tls}.
Clients connecting to a \sys instance attest the instance by verifying that it has a certificate issued by the \sys \gls{ca}.
A client must also provide a client certificate, which is stored upon creation of the security policy.
All further accesses (\ie, read, update, delete) to this policy are limited to the clients with the same certificate, and also require approval by the policy board.
Multiple clients can easily share the same certificate by running as part of a single security policy.

 \section{Evaluation}
\label{sec:evaluation}

We evaluate \sys with respect to its security and its performance.
First, we introduce a security analysis. 
Second, we measure the overheads both at the  micro-level in controlled environments and at the macro-level in real deployments.  

\subsection{Security Analysis}
\label{sec:security}

In terms of secret management, \sys provides stronger security guarantees compared to previous systems like Barbican~\cite{Barbican} or Vault~\cite{Vault}.
\sys protects against \emph{eavesdropping on any communication} by only supporting \gls{tls}-based communication using ciphers with \emph{perfect forward secrecy}~\cite{gunther1989identity}.
To avoid man-in-the-middle attacks by potentially compromised root \glspl{ca}~\cite{soghoian2011certified,huang2014analyzing}, we run our own \sys \gls{ca} to generate the \sys \gls{tls} certificates.
This CA runs inside a \gls{tee} and is controlled by a policy board to protect against malicious software updates.
Each \sys instance serves its \gls{ias} reports as a second way to be attested by its clients.
Clients are free to combine both approaches.  

\sys protects the \gls{cif} of data by encrypting all data and tracking the freshness if not otherwise ensured---for all data at rest, in transit or in main memory.
This requires us to effectively protect the secrets, such as the symmetric and/or asymmetric keys. 
\sys protects against unauthorized accesses to secrets by enforcing the following:
\begin{enumerate*}[label=\emph{(\roman*)}]
\item secrets are defined in the context of a security policy;
\item each application runs in the context of a single security policy;
\item only applications running in the context of a security policy are permitted to retrieve secrets of this security policy: the security policy specifies for each of its applications which secrets it is permitted to access; and
\item both the application code and its file system state are specified in the security policy and attested before the application can gain access to any secrets.
\end{enumerate*}

Access to security policies is controlled with the help of certificates.
Each client has to be authorized to access a security policy: a client must know the private key that corresponds to the public key used to create the security policy.
Note that no other entity, like the provider managing the \sys instance, can access this policy without knowing this private key. 
Therefore, only the client that creates a security policy controls access to this security policy. 
Any policy access must additionally be authorized by its policy board to protect against authorized but Byzantine client accesses.

\sys protects the \emph{confidentiality of stored secrets} by encrypting all secrets at rest with a randomly selected key only known to itself.
By always executing inside of a \gls{tee}, \sys protects against \emph{memory analysis of a running instance}.
A \sys instance is protected against some types of \emph{control of its storage backend} by providing:
\begin{enumerate*}[label=\emph{(\roman*)}]
\item protection against \emph{manipulation} (including rollbacks), by maintaining a Merkle tree of its files and storing this hash value in a sealed file (protection against rollbacks is ensured with the help of a monotonic counter); and
\item protection of the \emph{availability and durability of the storage backend} by the use of a trusted object storage like \textsc{Pesos}~\cite{PESOS:2018}.
\end{enumerate*}

\begin{figure}[t!]
    \includegraphics[scale=0.62]{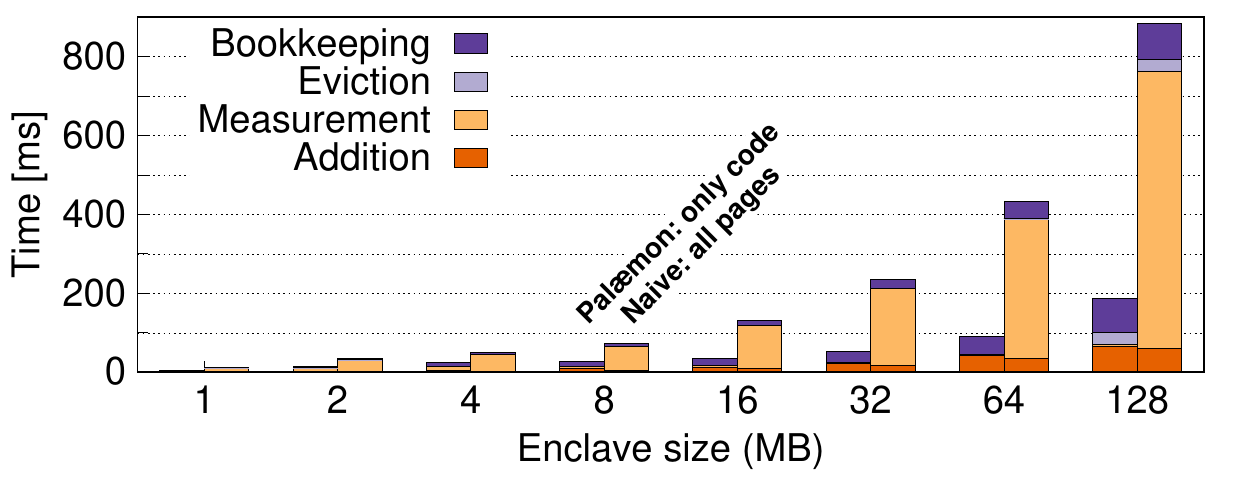}
    \caption{Startup times (80\unit{kB} binary) are typically not dominated by measurement overhead as \sys only measures code (left bars).}
    \label{fig:small-enclaves-startup}
    \spaceafterfloat
\end{figure}

\sys protects against \emph{attackers with superuser access} by executing securely inside a \gls{tee}.
All communication, files and data outside the \gls{tee} are always encrypted.
Hence, even users with superuser privileges cannot access or modify any secrets.
Access rules are defined and enforced per security policy, and a security policy can only be modified when an authorized client requests a change that must then be approved by the policy board of the security policy.

Although side-channel attacks are out of scope of this work, it is worth to mention that the underlying SCONE platform can protect against L1-based side channels attacks~\cite{oleksenko2018varys} and is hardened against Iago attacks~\cite{checkoway2013iago}.
To mitigate the various variants of Spectre~\cite{kocher2018spectre}, we can use LLVM-extensions, e.g., \emph{speculative load hardening}~\cite{SpecLH2019} that prevent exploitable speculation.

\subsection{Micro-benchmarks}

\textbf{Evaluation Settings.}
All our experiments are executed on a rack-based cluster of Dell PowerEdge R330 servers.
Each machine is equipped with an Intel Xeon E3-1270\,v6 CPU and 64\unit{GB} of RAM.
The machines are connected to a 20\unit{Gb/s} switched network.
\gls{sgx} is statically configured to reserve 128\unit{MB} of RAM for the \gls{epc}~\cite{costan2016intel}.
We use Ubuntu 16.04\,LTS with Linux kernel v4.13.0-38.
The CPUs use the latest microcode patch level.

The underlying SCONE runtime also supports \gls{emu} to run legacy applications without any \gls{tee} support.
We use \gls{emu} during the evaluation where indicated to highlight the performance overhead of the \gls{tee}.

\smallskip\noindent
\textbf{Enclave Startup Times.}
First, we evaluate how long it takes to start an application inside an \gls{sgx} enclave.
\sys automatically loads an application inside of an enclave with the help of a modified loader.
Setting up an enclave includes:
\begin{enumerate*}[label=\emph{(\roman*)}]
\item adding pages to the enclave;
\item measuring their content;
\item evicting pages if the enclave's size exceeds the \gls{epc}; and
\item bookkeeping tasks such as allocating memory and copying data.
\end{enumerate*}
We measured the throughput of each component with a micro-benchmark (see \autoref{tab:throughput}). 

On new \gls{sgx}-capable CPUs, the \sys runtime will dynamically allocate heap memory, \ie, startup times are mainly determined by the time it takes to load the code with some minimum heap.
When the runtime fails to allocate memory, it tries to add new heap pages to the enclave.
Current CPUs support \gls{sgx} enclave sizes of up to 64\unit{GB}.
This limit is expected to increase much further, \ie, we could run most applications inside of enclaves.

To ensure the integrity of an enclave, we need to measure all its code and initialized data segments.
The internal memory allocator is aware of the position of the contiguous enclave memory.
It will not use memory outside of the enclave, and it overwrites requested memory with zeros, avoiding measurement of added heap pages.
For small initial enclaves---which we expect to be common when adding heap memory dynamically---bookkeeping and page addition times are typically the dominant factors, despite the slow measuring speed (see \autoref{fig:small-enclaves-startup}). 

\smallskip\noindent
\textbf{Attestation and Configuration.}
\label{sec:measure-cas}
First, we evaluate how long it takes to attest and configure an application.
The advantage of \sys over the traditional way using \gls{ias} to perform attestation is that \sys runs on the local cluster. 
We measured the time it takes to perform the individual steps of remote attestation (\autoref{sec:attestation}).
The \gls{ias} experiment ran on servers in Europe and in Portland, OR, USA (close to \gls{ias} servers).
In the future, we will support both \gls{ias} and DCAP~\cite{intel-dc-attestation-primitives}.
\sys's attestation infrastructure will stay the same, as it attests other factors like the file system state. 

\begin{table}[t!]
\small
\caption{The average throughput of measuring pages is about an order of magnitude slower than evicting or adding pages.}
\vspace{-8pt}
\label{tab:throughput}
\setlength{\tabcolsep}{3pt}
\center
\rowcolors{1}{gray!10}{gray!0}
\begin{tabular}{>{\centering\arraybackslash}p{0.225\columnwidth}>{\centering\arraybackslash}p{0.225\columnwidth}>{\centering\arraybackslash}p{0.225\columnwidth}>{\centering\arraybackslash}p{0.225\columnwidth}}
  \rowcolor{gray!25}
  \textbf{Bookkeeping} & \textbf{Eviction} & \textbf{Measurement} & \textbf{Addition}  \\
\hline
1,292\unit{MB/s} & 1,219\unit{MB/s} & 148\unit{MB/s} & 2,853\unit{MB/s} \\
\hline

\hline
\end{tabular}
\end{table}

\begin{figure}[t!]
    \centering
    \includegraphics[scale=0.62]{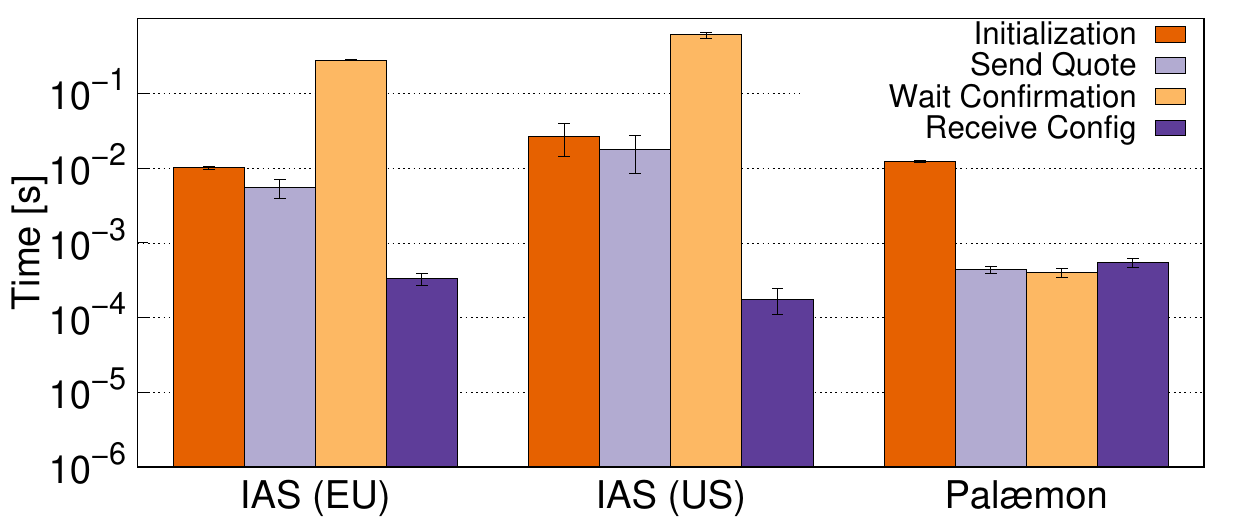}
    \caption{Attestation and configuration latencies: even when located close to Intel's \gls{ias} server, attestation with \gls{ias} takes about an order of magnitude longer than with \sys.}
    \label{fig:cas-latency}
    \spaceafterfloat
\end{figure}

\autoref{fig:cas-latency} shows the time it takes to:
\begin{enumerate*}[label=\emph{(\roman*)}]
\item initialize the necessary resources;
\item send the quote to \sys;
\item wait for \sys to confirm the successful attestation; and
\item receive the configuration.
\end{enumerate*}
The initialization phase includes key pair generation, DNS resolution, connection establishment, and \gls{tls} handshake with \sys.
Overall, the initialization time is similar for each attestation service and is dominated by the \gls{tls} handshake.

Obtaining and sending the quote takes longer for \gls{ias} variants for two reasons.
First, performing \gls{ias} attestation requires providing information that is embedded into the generated quote, which adds one round trip.
Second, \sys attestation cryptography (Ed25519~\cite{ed25519}) is less expensive than the one used by \gls{ias} (EPID~\cite{epid}).
However, the dominating factor for \gls{ias} is the time spent waiting for the attestation.
\sys has to verify the quote either by querying the \gls{ias} or by verifying the signature and looking up the public key of the \gls{qe}.
Overall, \sys attestation takes around 15\unit{ms} to complete, which is an order of magnitude faster than IAS attestation which takes 280\unit{ms} when performed from the USA, or 295\unit{ms} from Europe.

\begin{figure}[t!]
   \centering
  \includegraphics[scale=0.62]{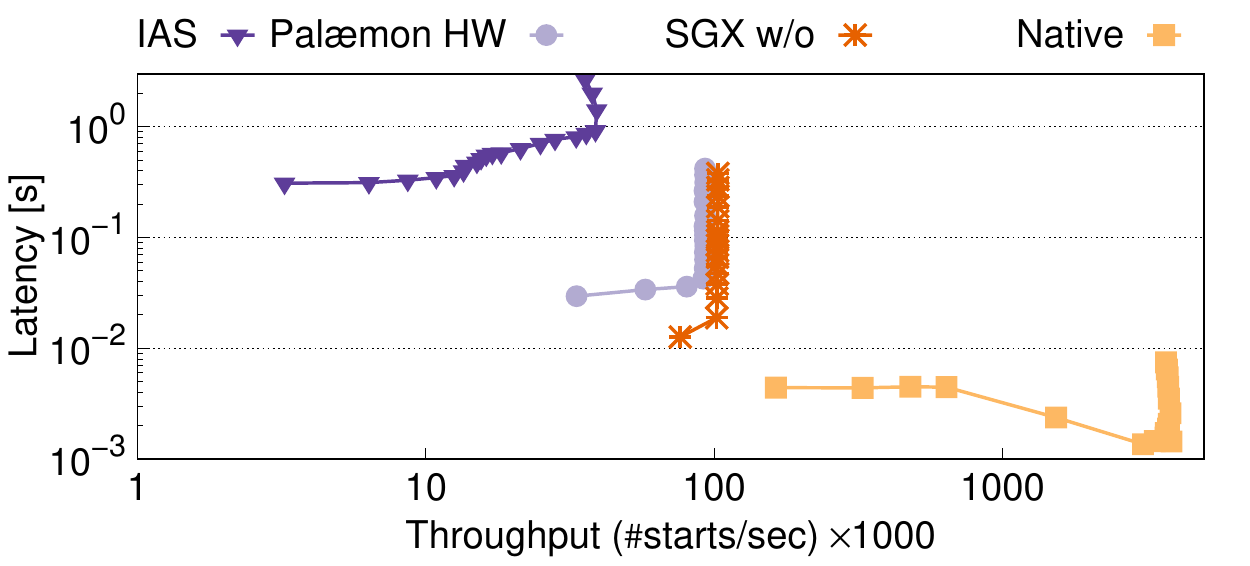}
   \caption{Startup latency and throughput using attestation variants.}
   \label{fig:start-throughout}
   \spaceafterfloat
\end{figure}

\sys also decouples application startup from \gls{ias}.
Our benchmark starts multiple minimal programs in parallel to measure the startup throughput and latency.
\autoref{fig:start-throughout} depicts the latency and throughput for different attestation variants.
In the \setup{Native} case (\gls{sgx} and attestation are not involved), the throughput scales well until all eight hyper-threads are fully utilized.
At this point, the system runs around 3,700 programs every second.
If the program is compiled with \gls{sgx} but without attestation (\setup{SGX w/o}), the throughput drops to about 100 executions per second.
This variant does not scale well with increasing parallelism.
We tracked down the bottleneck to the Intel \gls{sgx} driver synchronising \gls{epc} page (de)allocations with a single lock.
Since every enclave has to obtain \gls{epc} pages at roughly the same time, this lock basically enforces page requests to be served sequentially.

With \setup{IAS} and \setup{\sys}, the startup routine performs remote attestation before executing the actual program.
With \sys attestation, we quickly reach the maximal achievable start rate of about 90 runs per second.
IAS attestation needs a considerable amount of parallelism to partially hide the higher latency, reaching about 40 runs per second (60 parallel instances) at 1.4\unit{s} latency.

\begin{figure}[t!]
	\centering
	\includegraphics[scale=0.62]{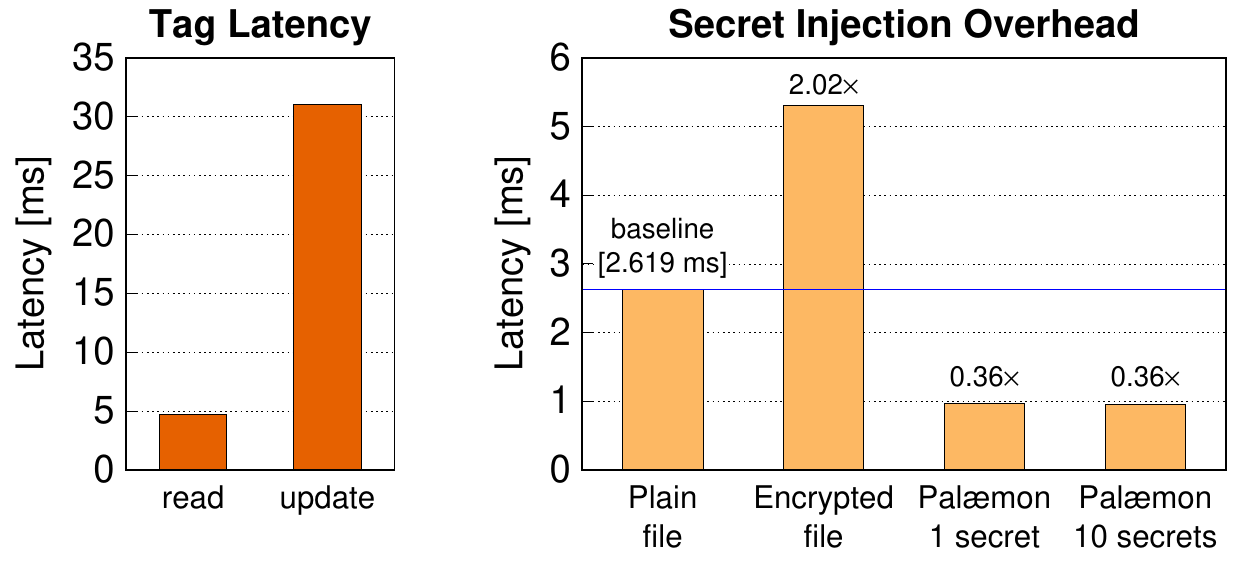}
	\caption{Left: latency of \sys tag reads and updates. Right: reading overhead for a file with 1 or 10 secrets normalized by the time to read a plain file.}
	\label{fig:microbench:tagreadupdate}
    \spaceafterfloat
    \vspace{-4pt}    
\end{figure}

\smallskip\noindent
\textbf{Rollback Protection.}
\sys protects against rollback attacks by ensuring that the root tags of all volumes of a process are sent to \sys on each file system synchronisation, file closing, as well as on program exits.
\sys stores the tags in its encrypted database. 
We measure the latency of the \sys runtime reading and updating the most recent tag in the \sys service to evaluate the overhead of rollback protection (\autoref{fig:microbench:tagreadupdate} left).
The update latency is roughly 6\,$\times$ higher than the read latency, as the \sys service database needs to be committed to disk for updates but not for reads.

\begin{figure}[t!]
    \includegraphics[scale=0.62,angle=270]{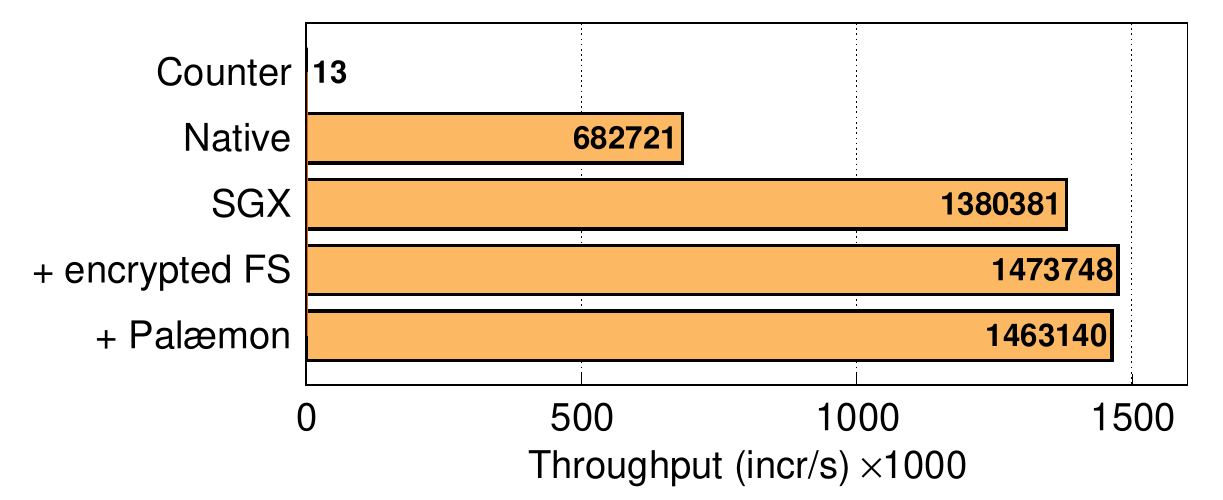}
	\caption{File-based monotonic counters are 5 orders of magnitude faster than platform counters for the same rollback protection.}
    \label{fig:amcs}
    \spaceafterfloat
\end{figure}

\sys itself is protected against rollbacks with the help of a monotonic counter.
To that end, we use the monotonic counters~\cite{SGXMonotonicCounters} provided by the \gls{sgx} platform. 
Independent measurements have shown that these counters allow between 4~\cite{ROTE} and 17~\cite{8023119} increments per second. 
TPM-based counters have a throughput of approximately 10 increments per second and wear out after 300\unit{k} to 1.4\unit{M} writes~\cite{Ariadne}.
The ROTE system~\cite{ROTE} stores the monotonic counters in memory of a group of servers, achieving a throughput of about 500 operations per second with 4 servers in a local area network.
However, the protection of ROTE against rollbacks is considered to be less robust than when using the platform counters.
In contrast, the anti-rollback protection offered by \sys in \emph{strict mode} is as safe as the underlying monotonic counters, \ie, \sys and its applications can only be rolled back if an attacker can roll back the monotonic counters of the platform.

We measure how fast we can increment a monotonic counter in the following scenarios (see \autoref{fig:amcs}):
\begin{enumerate*}[label=\emph{(\alph*)}]
\item using counters provided by the underlying platform and the Intel \gls{sgx} SDK;
\item \label{item:file-native} by opening a file, incrementing the integer stored in the file, writing back the new counter value and closing the file upon exit, when running in native mode;
\item \label{item:file-sgx} like \ref{item:file-native} but running inside \gls{sgx} enclaves without encrypting the file;
\item \label{item:file-encrypt} like \ref{item:file-sgx} but encrypting the file transparently with \sys; and
\item like \ref{item:file-encrypt} but also updating the tags with \sys by running in \emph{strict mode}, \ie, the file
is protected against rollbacks. 
\end{enumerate*}
Applications using \sys often do not need to use monotonic counters since files are rollback-protected. 
Still, they sometimes use them to track for instance the number of executions.

Using the platform's monotonic counters, we reach only  13 increments per seconds.
Note also that this approach requires applications to be rewritten in order to be protected against replay attacks.
Variant \ref{item:file-native} shows that we can reach a much higher throughput of 682\unit{k} increments per second when using a simple file-based counter.
When running inside of enclaves, throughput increases as files are transparently memory-mapped by the SCONE runtime to counter \gls{sgx} overhead.
When encrypting files, \sys automatically performs caching, increasing the throughput even more.
Sending the updated tag to a \sys instance only slightly reduces the throughput.
Our measurements show that applications could use a file-based counter and achieve throughputs that are 5 orders of magnitude higher than using the counters provided by the platform.
This approach relies on our assumption that system crashes are considered attacks.

\begin{figure}[t!]
	\centering
	\includegraphics[scale=0.62]{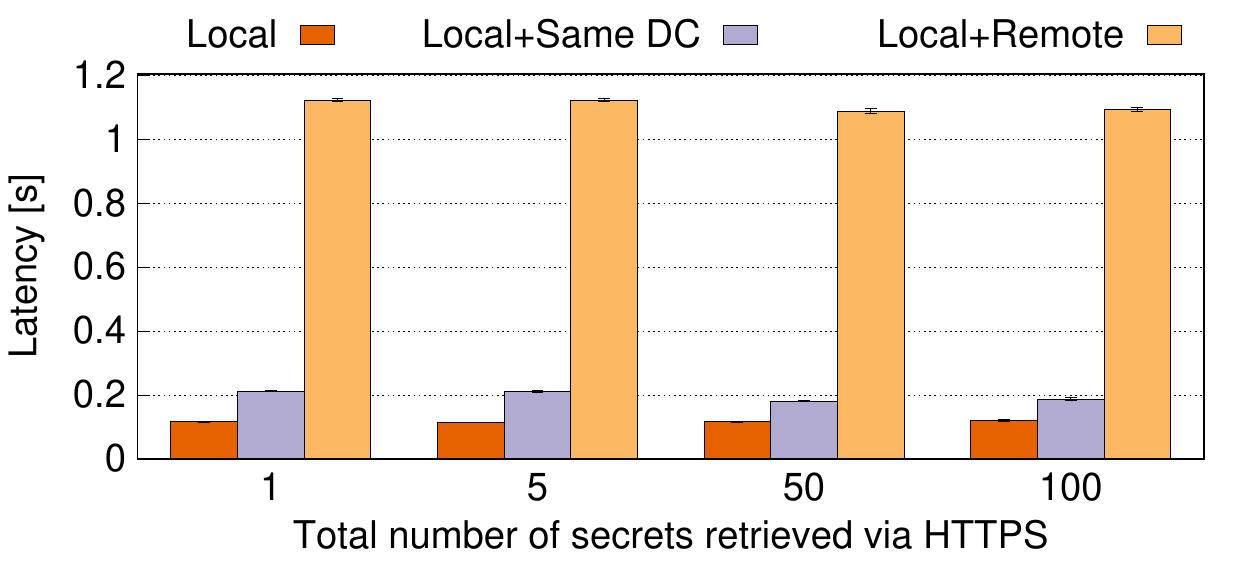}
	\caption{Latency to retrieve multiple secrets (up to 100) from a \sys service deployed locally, from the same data centre (DC) or from an instance running on a different continent.}
	\label{fig:microbench:pal:latency_fetch_secrets}
    \spaceafterfloat
\end{figure}

\smallskip\noindent
\textbf{Secret Injection Latency.}
We measure the impact of injecting secrets in a file by an application running inside an enclave.
To that end, we read a $4$\unit{kB} file in which we inject 1 and 10 secrets (\autoref{fig:microbench:tagreadupdate} right).
We show the latency as well as the overhead compared to the baseline on top of each bar.
\sys achieves better latencies for files with injected secrets---even compared to the \emph{plain file} baseline---because the secrets are injected during startup and stay in enclave memory. 

\smallskip\noindent
\textbf{Secret Access Latency.}
\sys supports the retrieval of keys from remote \sys services.
We measure the overhead of retrieving local and remote secrets, \ie, when using \sys in a decentralized fashion (\autoref{fig:microbench:pal:latency_fetch_secrets}).
There is no visible increase in latency when retrieving 1, 5, 50 and 100 keys of 32\unit{bytes}.
As a matter of fact, retrieving 50 or 100 keys consistently outperforms 1 or 5 keys. 
However, there is an impact if a peer service is located on a different continent instead of the same data centre.
This is mainly caused by the time it takes to establish of a \gls{tls} connection.

\begin{figure}[t!]
	\centering
	\includegraphics[scale=0.62]{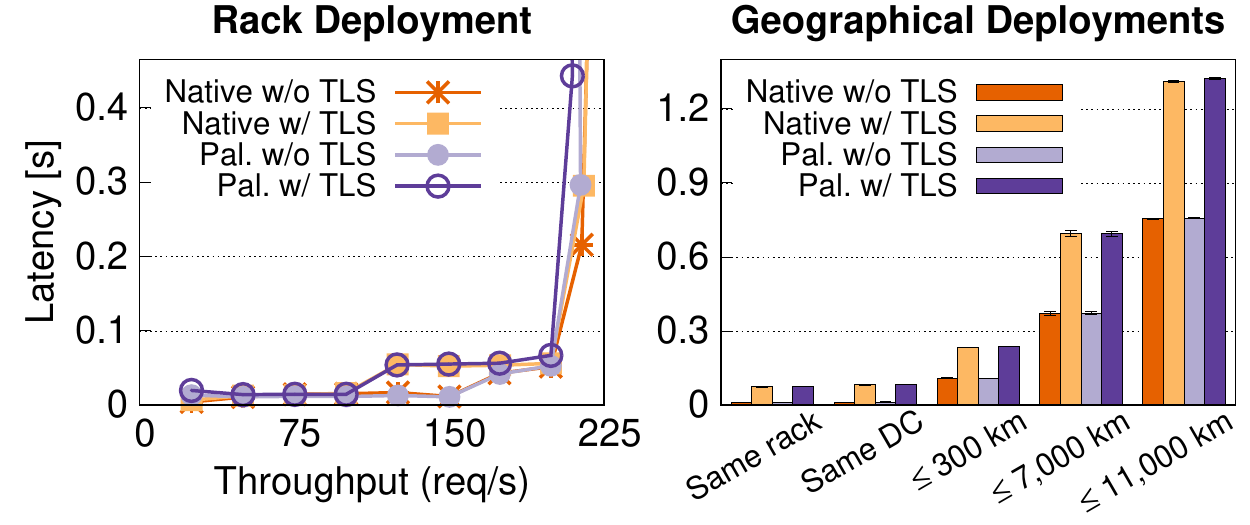}
	\caption{\sys's approval service: throughput/latency (left) and response latency (left) for different geographical deployments (from local to intercontinental).}
	\label{fig:microbench:approval}
    \spaceafterfloat
\end{figure}

\smallskip\noindent
\textbf{Approval Service.}
We measure the performance of the approval service running inside a \gls{tee} and compare it against a native version.
In both variants, we consider HTTP connections with and without \gls{tls} to show its impact.
The approval service and the client issuing requests run on the same rack.
We show the measured throughput/latency plots for these four combinations in \autoref{fig:microbench:approval} (left).
In these experiments, we issue approval requests at fixed rates (achieved throughput on the horizontal axis) until the response latencies spike. 
We observe that the \sys runtime over \gls{tls} achieves around 210 requests per second before the reply latencies spike. 
We consider these results satisfactory for our settings, where policy updates occur at slower rates.

Next, we investigate the impact on the response latency when the clients of the approval service are geographically distant. 
We configure this experiment with five different distances (see the horizontal axis of \autoref{fig:microbench:approval} (right)), from the closest (same rack) to the furthermost deployment with intercontinental latencies. 
We show the average latency of \sys's approval service response with the 95\% confidence interval.
As expected, the network latencies dominate the costs, with up to 1.36\unit{s} response time in the worst case. 

\subsection{Macro-benchmarks}
\hyphenation{MariaDB}
Our macro-benchmarks run using real-world systems, such as Barbican and Vault \glspl{kms}, the NGINX web-server, the Memcached cache system, the MariaDB database server (a fork of MySQL), and the ZooKeeper distributed coordination service.
Software versions used in macro-benchmarks are presented in \autoref{tab:apps}.
All these systems benefit from \sys for additional security guarantees; we evaluate its impact on performance.

\begin{figure}[t!]
	\centering
	\includegraphics[scale=0.62]{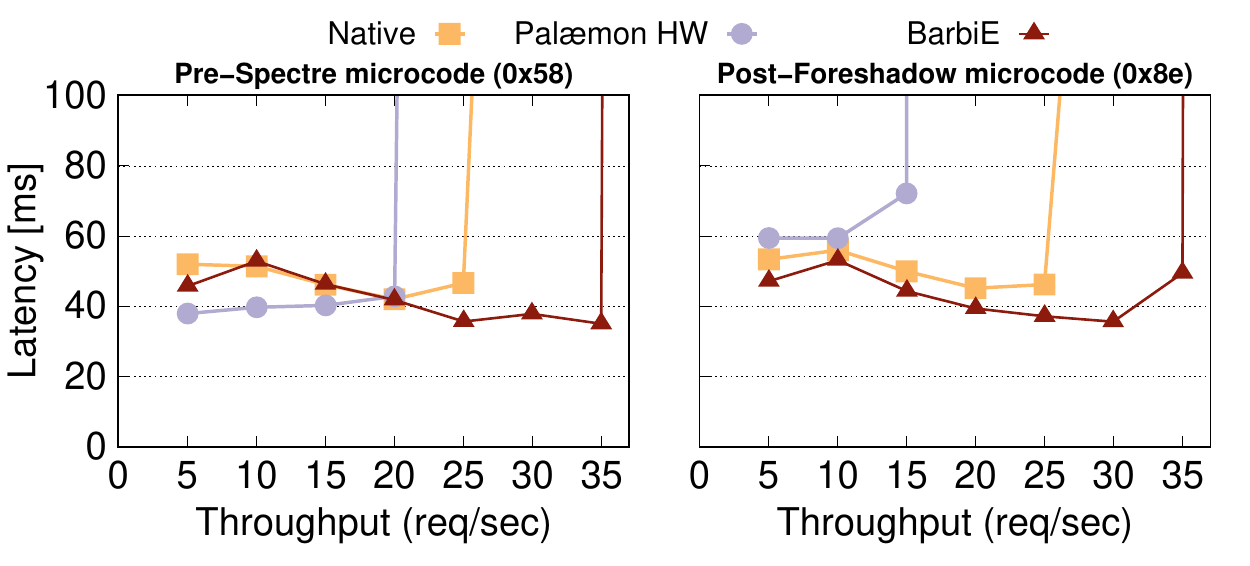}
	\caption{\emph{Barbican}. Throughput/latency of several variants (native, \sys and BarbiE~\cite{chakrabarti2017intel}) with two different microcodes.}
	\label{fig:microbench:Barbican:throughput}
    \spaceafterfloat
\end{figure}

\smallskip\noindent
\textbf{Barbican.}
We begin by measuring the throughput/latency ratio with different variants of Barbican (v5.0). 
\autoref{fig:microbench:Barbican:throughput} shows:
\begin{enumerate*}[label=\emph{(\roman*)}]
	\item native using a simple crypto plugin;
	\item \sys on \gls{sgx} hardware (HW); and
	\item BarbiE~\cite{chakrabarti2017intel} (Barbican using \gls{sgx} SDK v2.0 as HSM).
\end{enumerate*}
All variants run on CPython v2.7.14.
Additionally, we perform the Barbican measurements for CPUs with \emph{pre-Spectre} (version \texttt{0x58}) as well as \emph{post-Foreshadow} microcodes (\texttt{0x8e}).
\sys exhibits some overhead since the arguments of system calls must be checked by the \emph{syscall shield} while arguments are being copied out of the enclave and return values are copied back in.
BarbiE performs better than Barbican native due to its small \gls{tcb} and more efficient compiled code, rather than interpreted. 
Finally, the observed performance drop is of approximately 30\% when using the newer microcode.
We attribute this to the flushing of L1 cache on enclave exit,  required to mitigate the L1TF vulnerability, as also reported by Intel~\cite{intel2018l1tf} and \emph{Weichbrodt} \etal~\cite{weichbrodt2018sgxperf}.
BarbiE does not suffer as much since it requires less \gls{epc} paging and has a low number of enclave exits, in line with its number of requests per second.

\begin{figure}[t!]
  \centering
  \includegraphics[scale=0.62]{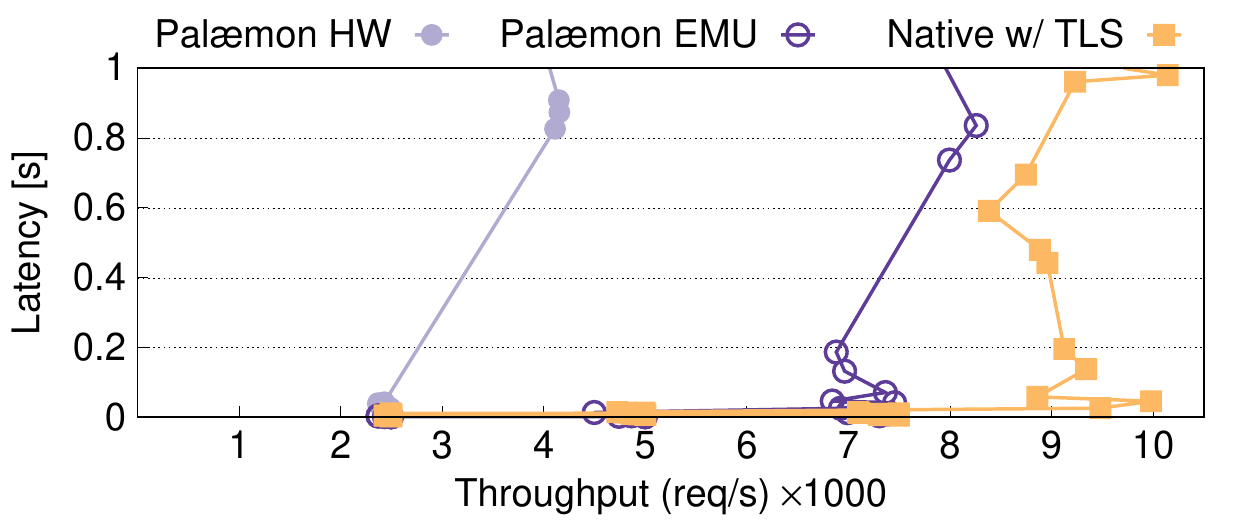}
  \caption{\emph{Vault.} Throughput/latency of native (with \gls{tls}), \sys in emulation and \sys in hardware mode.}
  \label{fig:microbench:vault:vault-palaemon-throughput-latency}
  \vspace{-4pt}
\end{figure}

\smallskip\noindent
\textbf{Vault.}
We evaluate Vault (v0.8.1) compiled by \texttt{gccgo} in Alpine Linux (\autoref{fig:microbench:vault:vault-palaemon-throughput-latency}). 
We use \texttt{wrk2}~\cite{wrk2} to retrieve secrets from Vault by providing it with an appropriate token.
Vault requires a heap of least 1.9\unit{GB} to start, \ie, the enclave is much larger than the \gls{epc}, so paging takes place. 
Our evaluation shows for instance that, for latencies below 1\unit{s}, \sys still achieves 61\% of native throughput when running in hardware, and up to 82\% when running in \gls{emu} mode. 

\begin{figure}[t!]
	\centering
	\includegraphics[scale=0.62]{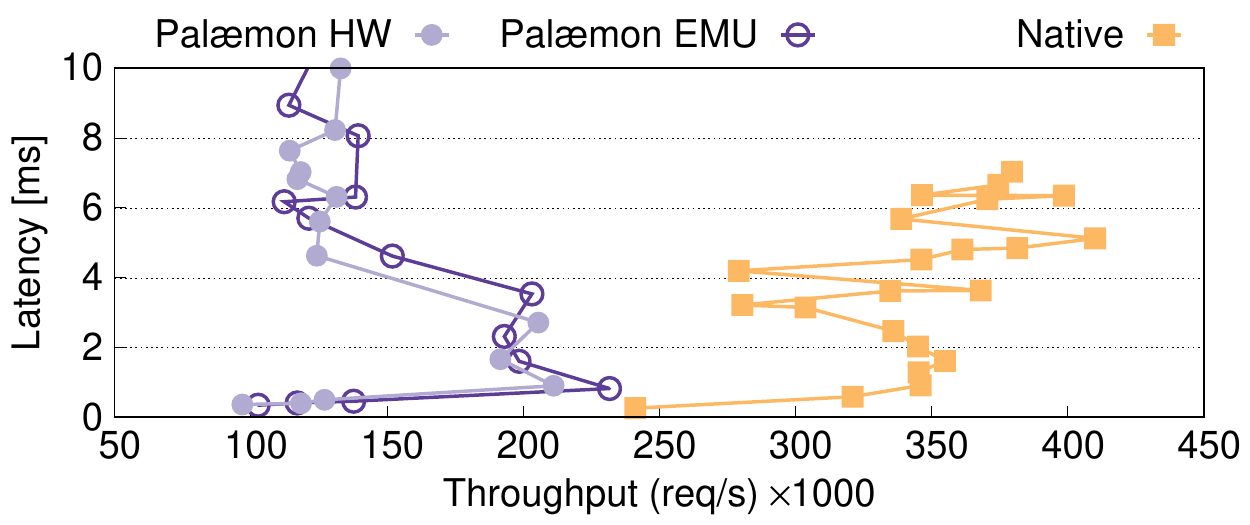}
	\caption{\emph{Memcached.} Throughput/latency of native, \sys in emulation and \sys in hardware mode.}
	\label{fig:macrobench:memcached}
    \spaceafterfloat
\end{figure}

\smallskip\noindent
\textbf{memcached.}
We evaluate the impact of \sys for running \gls{tls} protected \texttt{memcached}~\cite{fitzpatrick2004distributed}. 
In particular, \sys injects the certificates and private keys for \gls{tls} termination.
We use \texttt{memtier}~\cite{memtier} to load and stress \texttt{memcached}. 
\autoref{fig:macrobench:memcached} shows the measured latency and throughput of the evaluated systems.
We make a comparison between \sys and native \texttt{memcached} using \texttt{stunnel}~\cite{wong2001stunnel} \gls{tls} connections for both systems.
With latencies smaller than 3\unit{ms}, \sys achieves 59.5\% and 65.3\% of native throughput with hardware and \gls{emu} mode, respectively. 

\begin{figure*}[t!]
	\centering
	\includegraphics[scale=0.62]{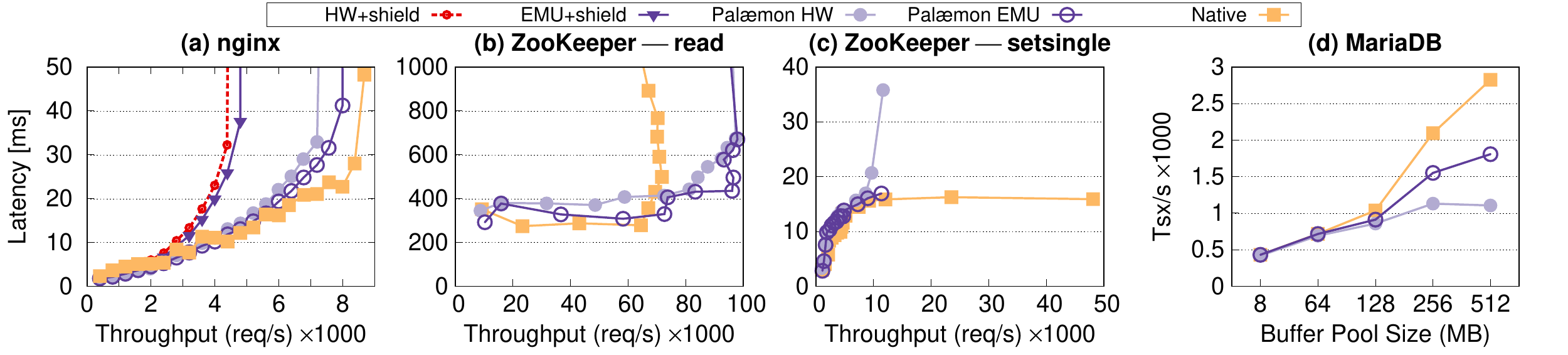}
	\caption{(a) Throughput/latency for GET requests on 67\unit{kB} files, in five variants of \texttt{nginx}.  ZooKeeper: read (b) and write (c) operations. (d) MariaDB with TPC-C benchmark (d): increasing buffer pool helps native more than \gls{emu} or hardware.}
	\label{fig:macrobench:apps}
    \spaceafterfloat
    \vspace{-2pt}
\end{figure*}

\smallskip\noindent
\textbf{NGINX.}
Along the same lines, we use an encrypted NGINX~\cite{reese2008nginx} container image and rely on \sys to:
\begin{enumerate*}[label=\emph{(\roman*)}]
\item encrypt all the files;
\item inject the certificates; and
\item inject private keys used by NGINX for \gls{tls} termination.
\end{enumerate*}
The benchmark issues \texttt{GET} requests on 67\unit{kB} files (nowadays' average size of an HTML web page~\cite{the-growth-of-web-page-size}) with the \texttt{wrk2} tool (see \autoref{fig:macrobench:apps}~(a)).
We see that the overhead of \gls{sgx} alone is less pronounced than that of encrypting all files. 
Tuning the caching done by NGINX could improve the performance when encrypting files. 
There is little difference between running in emulation mode and inside of an \gls{sgx} enclave, since not much paging is taking place.

\smallskip\noindent
\textbf{ZooKeeper.}
Next, we evaluate the overhead of \sys with the ZooKeeper coordination service.
We deploy a cluster of three nodes and evaluate three ZooKeeper variants:
\begin{enumerate*}[label=\emph{(\roman*)}]
	\item native using \texttt{stunnel}~\cite{wong2001stunnel} for \gls{tls} termination between servers;
	\item shielded ZooKeeper running together with the JVM in hardware mode; and
	\item \gls{emu} mode.
\end{enumerate*}
We use the ZooKeeper Benchmark~\cite{ZookeeperBench} to measure read and write throughput. 
The read throughput of the shielded versions is consistently better than the native one (\autoref{fig:macrobench:apps}~(b)).
The write throughput (\autoref{fig:macrobench:apps}~(c)) exhibits better performances in native mode, as it involves the execution of consensus~\cite{Hunt:2010:ZWC:1855840.1855851} via \gls{tls}, resulting in more code and system calls being executed. 
Our results are on par with \emph{SecureKeeper}~\cite{Brenner:2016:SCZ:2988336.2988350}, despite its use of an encryption proxy to protect the content only.  

\smallskip\noindent
\textbf{MariaDB.}
We conclude our macro-benchmarks by measuring the throughput of MariaDB configured to perform encryption at rest~\cite{mariadb-enc}.
We use \sys to inject a generated X.509 certificate, the private key and the encryption key.
We execute the TPC-C benchmark~\cite{council2010tpc} and vary the available buffer cache.
\autoref{fig:macrobench:apps}~(d) presents experimental results.
For small buffer pool sizes~\cite{bufferpool}, \ie, $<$128\unit{MB}, all configurations behave similarly since the main overhead is hardware I/O.
For larger buffer caches, \gls{epc} paging increases in hardware mode. 
Hence, adding more buffer cache reduces the throughput while it increases the throughput in emulation and native mode. 
A fair comparison with the recently proposed EnclaveDB~\cite{enclavedb-a-secure-database-using-sgx} system is currently not possible since it lacks paging support and its performance figures are only based on simulations.
 \section{Production Use Case}
\label{sec:usecase}

Finally, we describe a deployment of \sys in a real production environment for a company offering an online service for automatic conversion of handwritten documents into digital data via machine learning.
Typically, customers of this company want to acquire inference results and, due to the sensitive nature of the documents, ensure the confidentiality of the input images.
Additionally, the company wants to protect both the inference engine (implemented in Python) and its machine learning models. 

To achieve these security goals, the company has deployed \sys as follows. 
First, the company relies on SCONE's file system shields~\cite{arnautov2016scone} to encrypt Python code and models used for the inference.
The customers use the same mechanism to encrypt the input images. 
However, the company and the customers do not share with each other the keys and tags to decrypt and ensure the freshness of their digital assets.  
Instead, they define a dedicate security policy to define the access control to those.
Thereafter, they submit the policy to \sys after performing the attestation (see $\S$\ref{sec:attestation}) to ensure code integrity.
To process an image, it takes on average $~323ms$ and $~1202ms$ ($3.7\times$ slowdown) with the native and the \sys-enabled version, respectively. 
However, the result is less than $1.5$ seconds and thus considered acceptable in a production setting.
 \section{Related Work}
\label{sec:related}

\glsreset{kms}\Glspl{kms}~\cite{Barbican,ChefVault,Vault} provide an integrated approach for generating, managing and distributing cryptographic keys for devices and applications.
They are at the core of secure distributed systems and have been widely studied.
Many approaches rely on cryptographic techniques, often embedded in secure hardware modules~\cite{PKCS11,KMIP}.

Recent cloud computing frameworks integrate dedicated services for key management.
\emph{Barbican}~\cite{Barbican} and \emph{Vault}~\cite{Vault} are popular standalone \glspl{kms}.
Both rely on the \gls{os} for security, and thus consider a weaker threat model than \sys which considers attackers with superuser access (see \autoref{subsec:threat-model}).
\emph{Barbican} and \emph{Vault} can be protected against such attackers by running them on top of \sys; we evaluate these hardened variants in \autoref{sec:evaluation}.
Several major cloud providers also offer managed services to create and control encryption keys, \eg, Amazon~\cite{AmazonKMS}, Google~\cite{GoogleKMS} and Microsoft~\cite{MicrosoftKMS}.
Users must trust the providers to protect their secrets while \sys can be both managed by a provider and attested by users to establish trust in it.

While previous \glspl{kms} integrate \glspl{hsm} to provide better protection, we deem this approach vulnerable to the adversary \sys protects against.
An adversary with superuser access can eavesdrop on the \gls{hsm} to obtain secrets, or directly hijack the \gls{kms} and observe the secrets distributed to clients.

To the best of our knowledge, while \glspl{tee} have been widely used to secure many applications, only two systems are using \glspl{tee} to harden a \gls{kms} against adversaries with superuser privileges.
Researchers from Intel proposed the use of \gls{sgx} for securing \emph{Barbican}~\cite{chakrabarti2017intel}.
Along the same lines, Fortanix' \gls{sdkms}~\cite{Fortanix,Beekman:2017:CSA:3152701.3152710} also uses \gls{sgx} to securely generate, store and use cryptographic keys, certificates and various types of secrets.
Both approaches essentially provide a replacement for the functionality normally provided by \glspl{hsm} by using enclaves, hence reducing costs and providing better extensibility.
Like classical \glspl{hsm}, they still need passwords or \glspl{pin} in configuration files to authenticate clients.
In contrast, \sys provides an integrated approach to free itself from such sensitive identifiers and use the application code itself for authentication and authorization.
\sys also provides several additional features not found in other systems, \eg, advanced governance by a policy board, secret sharing between service instances and rollback protection.

To integrate \glspl{kms} into legacy applications without changing their source code, systems as Vault~\cite{Vault} process configuration files and environment variables using scripts (\eg, \texttt{consul-template} and \texttt{envconsul}) before executing the application~\cite{VaultDynamicSecrets}.
However, since this environment is maintained by the \gls{os}, it is thus accessible to attackers with superuser privileges.
Hence, this is not a viable solution to protect against privileged attackers.
\sys provides transparency by establishing the environment expected by the application inside the \gls{tee}, hence never exposing its secrets to the \gls{os}.

We finally mention that secrets are sometimes stored in configuration management services \emph{Chubby}~\cite{burrows2006chubby}, \emph{Consul}~\cite{Consul}, \emph{ZooKeeper}~\cite{Hunt:2010:ZWC:1855840.1855851} (or its SGX version \emph{SecureKeeper}~\cite{Brenner:2016:SCZ:2988336.2988350}) along with other configuration data.
Note that ZooKeeper neither encrypts data on disk nor does it protect its network communication.
With \sys, we can retrofit the necessary features to protect ZooKeeper in a cloud context (see \autoref{sec:evaluation}).
It is also worth mentioning that, \sys has been used not only in production, but also in several research works~\cite{clemmys, sgx-pyspark}.

 \section{Conclusion}
\label{sec:conclusion}

We introduced \sys, a service to manage trust in untrusted environments with Byzantine stakeholders. 
Unlike in the \glsdesc{bft} approach, we can enforce---via remote attestation---that the correct application code is executed.
In this way, we do not need to deploy multiple replicas to enforce integrity and freshness. Moreover, we also enforce confidentiality.
In order to support application updates, the root of trust of an application is a group of stakeholders---some of which might be Byzantine.
We protect applications with the help of \glspl{tee}: \sys clients can securely create secrets and protect access to these secrets with a security policy, even from insiders and attackers with superuser access.

To avoid source code changes, \sys passes the secrets to applications as arguments, environment variables and by transparently injecting these into files.
Our evaluation indicates that applications can achieve good throughput despite running in \glspl{tee}.
Throughput of our monotonic counters is 5 orders of magnitude higher than those offered by the SGX platform.
We will make \sys available to the research community.
 \section*{Acknowledgment}
The research leading to these results has received funding from the European Union's Horizon 2020 research and innovation programme and by the Swiss State Secretariat for Education, Research and Innovation (SERI) under the SecureCloud Project (\url{securecloudproject.eu}), grant agreement No~690111, as well as under the LEGaTO Project (\url{legato-project.eu}), grant agreement No~780681.

{
	\small
	\bibliographystyle{plain}
	\bibliography{references}

\begin{thebibliography}{10}

\bibitem{owens2010sqlite}
Grant Allen and Mike Owens.
\newblock {\em {The Definitive Guide to SQLite}}.
\newblock Apress, 2010.

\bibitem{AmazonKMS}
{Amazon.com, Inc.}
\newblock {AWS Key Management Service}.
\newblock \url{https://aws.amazon.com/kms/}, 2019.

\bibitem{anati2013innovative}
Ittai Anati, Shay Gueron, Simon Johnson, and Vincent Scarlata.
\newblock Innovative technology for cpu based attestation and sealing.
\newblock In {\em Proceedings of the 2nd International Workshop on Hardware and
  Architectural Support for Security and Privacy}, volume~13 of {\em HASP '13}.
  ACM, 2013.

\bibitem{arm-trustzone}
{ARM Limited}.
\newblock Building a secure system using trustzone technology.
\newblock White paper, 2009.

\bibitem{arnautov2016scone}
Sergei Arnautov, Bohdan Trach, Franz Gregor, Thomas Knauth, Andre Martin,
  Christian Priebe, Joshua Lind, Divya Muthukumaran, Dan
  O{\textquoteright}Keeffe, Mark~L. Stillwell, David Goltzsche, Dave Eyers,
  R{\"u}diger Kapitza, Peter Pietzuch, and Christof Fetzer.
\newblock {SCONE}: Secure linux containers with intel {SGX}.
\newblock In {\em 12th {USENIX} Symposium on Operating Systems Design and
  Implementation}, {OSDI} '16, pages 689--703. {USENIX} Association, 2016.

\bibitem{Beekman:2017:CSA:3152701.3152710}
Jethro~G. Beekman and Donald~E. Porter.
\newblock Challenges for scaling applications across enclaves.
\newblock In {\em Proceedings of the 2nd Workshop on System Software for
  Trusted Execution}, SysTEX'17, pages 8:1--8:2. ACM, 2017.

\bibitem{ed25519}
Daniel~J. Bernstein, Niels Duif, Tanja Lange, Peter Schwabe, and Bo-Yin Yang.
\newblock High-speed high-security signatures.
\newblock {\em Journal of Cryptographic Engineering}, 2(2):77--89, 2012.

\bibitem{8023119}
Marcus Brandenburger, Christian Cachin, Matthias Lorenz, and R\"udiger Kapitza.
\newblock Rollback and forking detection for trusted execution environments
  using lightweight collective memory.
\newblock In {\em 47th Annual IEEE/IFIP International Conference on Dependable
  Systems and Networks (DSN '17)}, pages 157--168, June 2017.

\bibitem{brasser2017software}
Ferdinand Brasser, Urs M{\"u}ller, Alexandra Dmitrienko, Kari Kostiainen,
  Srdjan Capkun, and Ahmad-Reza Sadeghi.
\newblock Software grand exposure: {SGX} cache attacks are practical.
\newblock In {\em 11th {USENIX} Workshop on Offensive Technologies}, {WOOT}
  '17. {USENIX} Association, 2017.

\bibitem{Brenner:2016:SCZ:2988336.2988350}
Stefan Brenner, Colin Wulf, David Goltzsche, Nico Weichbrodt, Matthias Lorenz,
  Christof Fetzer, Peter Pietzuch, and R\"{u}diger Kapitza.
\newblock Securekeeper: Confidential zookeeper using intel sgx.
\newblock In {\em Proceedings of the 17th International Middleware Conference},
  Middleware '16, pages 14:1--14:13. ACM, 2016.

\bibitem{epid}
E.~{Brickell} and J.~{Li}.
\newblock Enhanced privacy id from bilinear pairing for hardware authentication
  and attestation.
\newblock In {\em IEEE Second International Conference on Social Computing},
  SocialCom '10, pages 768--775, 2010.

\bibitem{burrows2006chubby}
Mike Burrows.
\newblock The chubby lock service for loosely-coupled distributed systems.
\newblock In {\em 7th {USENIX} Symposium on Operating Systems Design and
  Implementation ({OSDI} '06)}, pages 335--350. {USENIX} Association, 2006.

\bibitem{SpecLH2019}
Chanandler Carruth.
\newblock Speculative load hardening.
\newblock \url{https://llvm.org/docs/SpeculativeLoadHardening.html}, 2019.

\bibitem{castro1999practical}
Miguel Castro and Barbara Liskov.
\newblock Practical byzantine fault tolerance.
\newblock In {\em Proceedings of the Third Symposium on Operating Systems
  Design and Implementation}, OSDI '99, pages 173--186. USENIX Association,
  1999.

\bibitem{SGXMonotonicCounters}
Shanwei Cen and Bo~Zhang.
\newblock Trusted time and monotonic counters with intel software guard
  extensions platform services.
\newblock White paper, Intel Corporation, 2017.

\bibitem{chakrabarti2017intel}
Somnath {Chakrabarti}, Brandon {Baker}, and Mona {Vij}.
\newblock Intel sgx enabled key manager service with openstack barbican.
\newblock {\em arXiv e-prints}, 2017.

\bibitem{checkoway2013iago}
Stephen Checkoway and Hovav Shacham.
\newblock Iago attacks: Why the system call api is a bad untrusted rpc
  interface.
\newblock In {\em Proceedings of the Eighteenth International Conference on
  Architectural Support for Programming Languages and Operating Systems},
  ASPLOS '13, pages 253--264. ACM, 2013.

\bibitem{chen2018sgxpectre}
Guoxing {Chen}, Sanchuan {Chen}, Yuan {Xiao}, Yinqian {Zhang}, Zhiqiang {Lin},
  and Ten~H. {Lai}.
\newblock Sgxpectre attacks: Stealing intel secrets from sgx enclaves via
  speculative execution.
\newblock {\em arXiv e-prints}, 2018.

\bibitem{KCook}
Keen Cook.
\newblock The state of kernel self protection project.
\newblock In {\em Linux Security Summit}, 2016.
\newblock
  \url{https://www.linux.com/videos/state-kernel-self-protection-project-kees-cook-google}.

\bibitem{costan2016intel}
Victor Costan and Srinivas Devadas.
\newblock Intel sgx explained.
\newblock {\em IACR Cryptology ePrint Archive}, 2016(086):1--118, 2016.

\bibitem{VaultDynamicSecrets}
Armon Dadgar.
\newblock Why we need dynamic secrets.
\newblock \url{https://www.hashicorp.com/blog/why-we-need-dynamic-secrets},
  2018.

\bibitem{Snowden}
Christopher Drew and Somini Sengupta.
\newblock {NSA Leak Puts Focus on System Administrators}.
\newblock In {\em New York Times}, 2013.
\newblock
  \url{https://www.nytimes.com/2013/06/24/technology/nsa-leak-puts-focus-on-system-administrators.html}.

\bibitem{du2017secure}
Zhao-Hui {Du}, Zhiwei {Ying}, Zhenke {Ma}, Yufei {Mai}, Phoebe {Wang}, Jesse
  {Liu}, and Jesse {Fang}.
\newblock Secure encrypted virtualization is unsecure.
\newblock {\em arXiv e-prints}, 2017.

\bibitem{amazontradesecret}
{Emont, Jon and Stevens, Laura and McMillan, Robert}.
\newblock {Amazon Investigates Employees Leaking Data for Bribes}.
\newblock
  \url{https://www.wsj.com/articles/amazon-investigates-employees-leaking-data-for-bribes-1537106401},
  2018.

\bibitem{fitzpatrick2004distributed}
Brad Fitzpatrick.
\newblock {Distributed Caching with Memcached}.
\newblock {\em Linux Journal}, 2004.

\bibitem{NSAhuntSA}
Ryan Gallagher and Peter Maass.
\newblock {Inside the NSA's Secret Efforts to Hunt and Hack System
  Administrators}.
\newblock In {\em The Intercept\_}, 2014.
\newblock
  \url{https://theintercept.com/2014/03/20/inside-nsa-secret-efforts-hunt-hack-system-administrators/}.

\bibitem{Goldman:2006:LRA:1179474.1179481}
Kenneth Goldman, Ronald Perez, and Reiner Sailer.
\newblock Linking remote attestation to secure tunnel endpoints.
\newblock In {\em Proceedings of the First ACM Workshop on Scalable Trusted
  Computing}, STC '06, pages 21--24. ACM, 2006.

\bibitem{Rootin5secs}
Dan Goodin.
\newblock {``Most serious'' Linux privilege-escalation bug ever is under active
  exploit (updated)}.
\newblock In {\em Ars Technica}, 2016.
\newblock
  \url{https://arstechnica.com/information-technology/2016/10/most-serious-linux-privilege-escalation-bug-ever-is-under-active-exploit/}.

\bibitem{GoogleKMS}
{Google LLC}.
\newblock {Google Cloud Key Management Service}.
\newblock \url{https://cloud.google.com/kms/}, 2019.

\bibitem{sgx-migration}
Jinyu Gu, Zhichao Hua, Yubin Xia, Haibo Chen, Binyu Zang, Haibing Guan, and
  Jinming Li.
\newblock Secure live migration of sgx enclaves on untrusted cloud.
\newblock In {\em Proceedings of tge 47th Annual IEEE/IFIP International
  Conference on Dependable Systems and Networks (DSN)}, pages 225--236. IEEE,
  2017.

\bibitem{gunther1989identity}
Christoph~G. G{\"u}nther.
\newblock An identity-based key-exchange protocol.
\newblock In {\em Advances in Cryptology}, EUROCRYPT '89, pages 29--37.
  Springer Berlin Heidelberg, 1990.

\bibitem{Consul}
{HashiCorp}.
\newblock {Consul: Service Discovery and Configuration Management}.
\newblock \url{https://www.consul.io}, 2019.

\bibitem{Vault}
{HashiCorp}.
\newblock {Vault}.
\newblock \url{https://www.vaultproject.io/}, 2019.

\bibitem{hetzelt2017security}
Felicitas Hetzelt and Robert Buhren.
\newblock Security analysis of encrypted virtual machines.
\newblock In {\em Proceedings of the 13th ACM SIGPLAN/SIGOPS International
  Conference on Virtual Execution Environments}, VEE '17, pages 129--142. ACM,
  2017.

\bibitem{hoekstra2013using}
Matthew Hoekstra, Reshma Lal, Pradeep Pappachan, Vinay Phegade, and Juan
  Del~Cuvillo.
\newblock Using innovative instructions to create trustworthy software
  solutions.
\newblock In {\em Proceedings of the 2nd International Workshop on Hardware and
  Architectural Support for Security and Privacy}, HASP '13, pages 11:1--11:1.
  ACM, 2013.

\bibitem{huang2014analyzing}
L.~S. {Huang}, A.~{Rice}, E.~{Ellingsen}, and C.~{Jackson}.
\newblock Analyzing forged ssl certificates in the wild.
\newblock In {\em 2014 IEEE Symposium on Security and Privacy}, S\&P '14, pages
  83--97, 2014.

\bibitem{Hunt:2010:ZWC:1855840.1855851}
Patrick Hunt, Mahadev Konar, Flavio~P. Junqueira, and Benjamin Reed.
\newblock Zookeeper: Wait-free coordination for internet-scale systems.
\newblock In {\em 2010 {USENIX} Annual Technical Conference}, {USENIX} {ATC}
  '10. {USENIX} Association, 2010.

\bibitem{IASv3Spec}
{Intel Corporation}.
\newblock {Attestation Service for Intel Software Guard Extensions (Intel SGX):
  API Documentation}.
\newblock
  \url{https://software.intel.com/sites/default/files/managed/7e/3b/ias-api-spec.pdf},
  2018.

\bibitem{intel2018l1tf}
{Intel Corporation}.
\newblock Resources and response to side channel {L1TF}.
\newblock
  \url{https://www.intel.com/content/www/us/en/architecture-and-technology/l1tf.html},
  2018.

\bibitem{johnson2016intel}
Simon Johnson, Vinnie Scarlata, Carlos Rozas, Ernie Brickell, and Frank Mckeen.
\newblock {Intel Software Guard Extensions: EPID Provisioning and Attestation
  Services}.
\newblock In {\em Intel Whitepaper}, 2016.
\newblock
  \url{https://software.intel.com/en-us/blogs/2016/03/09/intel-sgx-epid-provisioning-and-attestation-services}.

\bibitem{the-growth-of-web-page-size}
{KeyCDN}.
\newblock The growth of web page size.
\newblock \url{https://www.keycdn.com/support/the-growth-of-web-page-size/},
  2017.

\bibitem{kocher2018spectre}
Paul Kocher, Jann Horn, Anders Fogh, , Daniel Genkin, Daniel Gruss, Werner
  Haas, Mike Hamburg, Moritz Lipp, Stefan Mangard, Thomas Prescher, Michael
  Schwarz, and Yuval Yarom.
\newblock {Spectre Attacks: Exploiting Speculative Execution}.
\newblock In {\em 40th IEEE Symposium on Security and Privacy (S\&P'19)}, 2019.

\bibitem{PESOS:2018}
Robert Krahn, Bohdan Trach, Anjo Vahldiek-Oberwagner, Thomas Knauth, Pramod
  Bhatotia, and Christof Fetzer.
\newblock Pesos: Policy enhanced secure object store.
\newblock In {\em Proceedings of the Thirteenth EuroSys Conference}, EuroSys
  '18, pages 25:1--25:17. ACM, 2018.

\bibitem{Fortanix}
Ambuj Kumar, Anand Kashyap, Vinay Phegade, and Jesse Schrater.
\newblock {Self-Defending Key Management Service with Intel Software Guard
  Extensions}.
\newblock White paper, Fortanix, 2018.

\bibitem{sgx-pyspark}
Do~Le~Quoc, Franz Gregor, Jatinder Singh, and Christof Fetzer.
\newblock {SGX-PySpark}: Secure distributed data analytics.
\newblock In {\em Proceedings of the World Wide Web Conference (WWW)}, 2019.

\bibitem{lee2017inferring}
Sangho Lee, Ming-Wei Shih, Prasun Gera, Taesoo Kim, Hyesoon Kim, and Marcus
  Peinado.
\newblock Inferring fine-grained control flow inside {SGX} enclaves with branch
  shadowing.
\newblock In {\em 26th {USENIX} Security Symposium}, {USENIX} Security '17,
  pages 557--574. {USENIX} Association, 2017.

\bibitem{ZookeeperBench}
Chen Liang, Andrew Ferguson, and Rodrigo Fonseca.
\newblock {Zookeeper Benchmark}.
\newblock \url{https://github.com/brownsys/zookeeper-benchmark}, 2014.

\bibitem{mariadb-enc}
{MariaDB}.
\newblock Data-at-rest encryption.
\newblock
  \url{https://mariadb.com/kb/en/library/data-at-rest-encryption-overview/},
  2019.

\bibitem{ROTE}
Sinisa Matetic, Mansoor Ahmed, Kari Kostiainen, Aritra Dhar, David Sommer,
  Arthur Gervais, Ari Juels, and Srdjan Capkun.
\newblock {ROTE}: Rollback protection for trusted execution.
\newblock In {\em 26th {USENIX} Security Symposium}, {USENIX} Security '17,
  pages 1289--1306. {USENIX} Association, 2017.

\bibitem{Matsakis:2014:RL:2692956.2663188}
Nicholas~D. Matsakis and Felix~S. Klock, II.
\newblock The rust language.
\newblock In {\em Proceedings of the 2014 ACM SIGAda Annual Conference on High
  Integrity Language Technology}, HILT '14, pages 103--104. ACM, 2014.

\bibitem{MicrosoftKMS}
{Microsoft Corporation}.
\newblock {Microsoft Azure Key Vault}.
\newblock \url{https://azure.microsoft.com/en-us/services/key-vault/}, 2019.

\bibitem{morbitzer2018severed}
Mathias Morbitzer, Manuel Huber, Julian Horsch, and Sascha Wessel.
\newblock Severed: Subverting amd's virtual machine encryption.
\newblock In {\em Proceedings of the 11th European Workshop on Systems
  Security}, EuroSec'18, pages 1:1--1:6. ACM, 2018.

\bibitem{ChefVault}
Kevin Moser, Eli Klein, Joey Geiger, Joshua Timberman, James FitzGibbon, and
  Thom May.
\newblock {Chef Vault}.
\newblock \url{https://github.com/chef/chef-vault}, 2019.

\bibitem{noor2013trust}
Talal~H. Noor, Quan~Z. Sheng, Sherali Zeadally, and Jian Yu.
\newblock Trust management of services in cloud environments: Obstacles and
  solutions.
\newblock {\em ACM Comput. Surv.}, 46(1):12:1--12:30, 2013.

\bibitem{KMIP}
{OASIS}.
\newblock {OASIS Key Management Interoperability Protocol (KMIP) TC}.
\newblock
  \url{https://www.oasis-open.org/committees/tc_home.php?wg_abbrev=kmip}, 2018.

\bibitem{PKCS11}
{OASIS}.
\newblock {OASIS PKCS 11 TC}.
\newblock
  \url{https://www.oasis-open.org/committees/tc_home.php?wg_abbrev=pkcs11},
  2018.

\bibitem{oleksenko2018varys}
Oleksii Oleksenko, Bohdan Trach, Robert Krahn, Mark Silberstein, and Christof
  Fetzer.
\newblock Varys: Protecting {SGX} enclaves from practical side-channel attacks.
\newblock In {\em 2018 {USENIX} Annual Technical Conference}, {USENIX} {ATC}
  '18, pages 227--240. {USENIX} Association, 2018.

\bibitem{Barbican}
{OpenStack}.
\newblock {Barbican}.
\newblock \url{https://wiki.openstack.org/wiki/Barbican}, 2018.

\bibitem{bufferpool}
{Oracle Corporation}.
\newblock {InnoDB Startup Options and System Variables, MySQL 8.0 Reference
  Manual}.
\newblock \url{https://dev.mysql.com/doc/refman/8.0/en/innodb-parameters.html},
  2019.

\bibitem{enclavedb-a-secure-database-using-sgx}
Christian Priebe, Kapil Vaswani, and Manuel Costa.
\newblock {EnclaveDB: A Secure Database Using SGX}.
\newblock In {\em 2018 IEEE Symposium on Security and Privacy}, S\&P '18, pages
  264--278, 2018.

\bibitem{memtier}
{Redis Labs}.
\newblock {{NoSQL Redis and Memcached traffic generation and benchmarking
  tool}}.
\newblock \url{https://github.com/RedisLabs/memtier_benchmark}, 2019.

\bibitem{reese2008nginx}
Will Reese.
\newblock Nginx: The high-performance web server and reverse proxy.
\newblock {\em Linux Journal}, 2008(173), 2008.

\bibitem{win8tradesecret}
{Reuters}.
\newblock {Ex-Microsoft employee charged with leaking trade secrets to
  blogger}.
\newblock
  \url{https://www.reuters.com/article/us-microsoft-tradesecret-idUSBREA2J07K20140320},
  2014.

\bibitem{intel-dc-attestation-primitives}
Vinnie Scarlata, Simon Johnson, James Beaney, and Piotr Zmijewski.
\newblock {Supporting Third Party Attestation for Intel SGX with Intel Data
  Center Attestation Primitives}.
\newblock White paper, Intel Corporation, 2018.

\bibitem{soghoian2011certified}
Christopher Soghoian and Sid Stamm.
\newblock Certified lies: Detecting and defeating government interception
  attacks against ssl (short paper).
\newblock In {\em Financial Cryptography and Data Security}, pages 250--259.
  Springer Berlin Heidelberg, 2012.

\bibitem{Ariadne}
Raoul Strackx and Frank Piessens.
\newblock Ariadne: A minimal approach to state continuity.
\newblock In {\em 25th {USENIX} Security Symposium}, {USENIX} Security '16,
  pages 875--892. {USENIX} Association, 2016.

\bibitem{wrk2}
Gil Tene et~al.
\newblock {wrk2 HTTP benchmarking tool}.
\newblock \url{https://github.com/giltene/wrk2}, 2018.

\bibitem{clemmys}
Bohdan Trach, Oleksii Oleksenko, Franz Gregor, Pramod Bhatotia, and Christof
  Fetzer.
\newblock Clemmys: Towards secure remote execution in {FaaS}.
\newblock In {\em Proceedings of the 12th ACM International Conference on
  Systems and Storage (SYSTOR)}, 2019.

\bibitem{council2010tpc}
{Transaction Processing Performance Council}.
\newblock {TPC Benchmark C}.
\newblock
  \url{http://www.tpc.org/tpc_documents_current_versions/pdf/tpc-c_v5.11.0.pdf},
  2010.

\bibitem{vanbulck2018foreshadow}
Jo~Van~Bulck, Marina Minkin, Ofir Weisse, Daniel Genkin, Baris Kasikci, Frank
  Piessens, Mark Silberstein, Thomas~F. Wenisch, Yuval Yarom, and Raoul
  Strackx.
\newblock Foreshadow: Extracting the keys to the intel {SGX} kingdom with
  transient out-of-order execution.
\newblock In {\em 27th {USENIX} Security Symposium}, {USENIX} Security '18,
  pages 991--1008. {USENIX} Association, 2018.
\newblock See also technical report
  Foreshadow-NG~\cite{weisse2018foreshadowNG}.

\bibitem{weichbrodt2018sgxperf}
Nico Weichbrodt, Pierre-Louis Aublin, and R\"{u}diger Kapitza.
\newblock sgx-perf: A performance analysis tool for intel sgx enclaves.
\newblock In {\em Proceedings of the 19th International Middleware Conference},
  Middleware '18, pages 201--213. ACM, 2018.

\bibitem{weisse2018foreshadowNG}
Ofir Weisse, Jo~Van~Bulck, Marina Minkin, Daniel Genkin, Baris Kasikci, Frank
  Piessens, Mark Silberstein, Raoul Strackx, Thomas~F. Wenisch, and Yuval
  Yarom.
\newblock {Foreshadow-NG}: Breaking the virtual memory abstraction with
  transient out-of-order execution.
\newblock Technical report, 2018.
\newblock See also {USENIX} Security paper
  Foreshadow~\cite{vanbulck2018foreshadow}.

\bibitem{wong2001stunnel}
Wesley Wong.
\newblock {Stunnel: SSLing Internet Services Easily}.
\newblock White paper, SANS Institute, 2001.

\bibitem{6681007}
Y.~{Yu}, H.~{Wang}, B.~{Liu}, and G.~{Yin}.
\newblock A trusted remote attestation model based on trusted computing.
\newblock In {\em 12th IEEE International Conference on Trust, Security and
  Privacy in Computing and Communications}, TrustCom '13, pages 1504--1509,
  2013.

\bibitem{NSAKeynote}
Kim Zetter.
\newblock {NSA Hacker Chief Explains How to Keep Him Out of Your System}.
\newblock In {\em Wired}, 2016.
\newblock
  \url{https://www.wired.com/2016/01/nsa-hacker-chief-explains-how-to-keep-him-out-of-your-system/}.

\end{thebibliography}
}

\end{document}